\documentclass[a4paper]{article}

\usepackage{INTERSPEECH2021}
\usepackage{graphicx}
\usepackage{amssymb, amsmath, bm, mathtools,array}
\usepackage{textcomp}
\usepackage{hyperref}
\usepackage{verbatim}
\usepackage{lipsum} 
\usepackage{multirow}
\usepackage{siunitx}
\usepackage{diagbox}
\usepackage{dsfont}
\usepackage{colortbl}
\usepackage{booktabs}
\usepackage{caption}
\usepackage{subcaption}

\newcommand{\bs}[1]{\boldsymbol{#1}}

\RequirePackage{color}\definecolor{BLUE}{rgb}{0,0,0}

\newcommand{\systemI}{\texttt{LCNN-trim-pad}}
\newcommand{\systemII}{\texttt{LCNN-attention}}
\newcommand{\systemIII}{\texttt{LCNN-LSTM-sum}}

\title{A Comparative Study on Recent Neural Spoofing Countermeasures for Synthetic Speech Detection}
\name{Xin Wang, Junich Yamagishi}
\address{National Institute of Informatics, Chiyoda-ku, Tokyo, Japan}
\email{wangxin@nii.ac.jp, jyamagis@nii.ac.jp}

\begin{document}

\maketitle
\begin{abstract}
A great deal of recent research effort on speech spoofing countermeasures has been invested into back-end neural networks and training criteria. We contribute to this effort with a comparative perspective in this study. Our comparison of countermeasure models on the ASVspoof 2019 logical access task takes into account recently proposed margin-based training criteria, widely used front ends, and common strategies to deal with varied-length input trials. We also measured intra-model differences through multiple training-evaluation rounds with random initialization. Our statistical analysis demonstrates that the performance of the same model may be significantly different when just changing the random initial seed. Thus, we recommend similar analysis or multiple training-evaluation rounds for further research on the database. Despite the intra-model differences, we observed a few promising techniques such as the average pooling to process varied-length inputs and a new hyper-parameter-free loss function. The two techniques led to the best single model in our experiment, which achieved an equal error rate of 1.92\% and was significantly different in statistical sense from most of the other experimental models.

\end{abstract}
\noindent\textbf{Index Terms}: anti-spoofing, countermeasure, ASVspoof 2019, logical access, deep learning, significance test

\section{Introduction}
Neural-network-based countermeasures (CMs) are a hot topic in speech anti-spoofing research \cite{Nautsch2021}.
Many studies have used neural networks and loss functions \cite{lavrentyeva2019stc,zhang2020one,Chen2020Odyssey} proposed for face verification and image classification tasks, for example, light convolution network (LCNN) \cite{wu2018light} and margin-based softmax \cite{wang2018cosface,wang2018additive}. 
There is, however, a potential glitch of using an image-oriented neural network for speech because speech trials have varied length, while images do not. 
No consensus has been reached: some studies trim or pad input speech trials while others squeeze them through pooling or attention.
On loss functions, although the margin-based softmax is well supported by empirical studies in the image field, comparative studies for speech anti-spoofing are lacking. Whether or not the margin-based softmax is well worth the efforts to search for a good hyper-parameter set for the margin is unknown.  

We have two purposes in this study. First, we summarize and compare the strategies to deal with varied-length input and a few loss functions reported in the recent speech anti-spoofing literatures. Second, we introduce a simple hyper-parameter-free loss function. We built a few CMs by combining the aforementioned components and compared their performance on the ASVspoof2019 logical access (LA) database across front ends based on linear frequency cepstral coefficients (LFCCs), linear filter bank coefficients (LFBs), and spectrograms.

Through training and evaluating each CM multiple times with random initialization, we observed that the performance variation through multiple runs can be more statistically discernible than inter-model differences. This intra-model variation calls for more caution when reporting and interpreting the CM performance on the database. We suggest conducting statistical analysis or simply multiple runs for future work.

Despite the intra-model variation, our results suggest that CMs can efficiently process the varied-length speech trials using attention or simply average pooling. Results on loss functions demonstrate the decent performance of a simple sigmoid function, which makes the gain of the margin-based softmax trivial. 
The new and simple loss function based on P2SGrad \cite{zhang2019p2sgrad} is a competitive alternative. By combining the new loss function and an LCNN with average pooling, one of the CM achieved an equal error rate (EER) of 1.92\% in the best training-evaluation round. This is one of the lowest EERs among existing single CMs on the database without data augmentation \cite{zhang2020one, Chen2020Odyssey, luo2021capsule, li2021replay, tak2020end}, even though the differences may not be statistically significant.

This paper briefly describes recent neural CMs in Sec.~\ref{sec:overview} and introduces the P2SGrad-based loss function. It then details the experimental design and results in Sec.~\ref{seq:exp}. 
Note that this work focuses on the CM back end, a better CM also requires joint efforts from the front end \cite{Tak2020,Chettri2020} and model ensemble \cite{Nautsch2021,Tak2020is}.

\section{Brief Overview of Neural CMs}
\label{sec:overview}
We focus on CMs that convert a sequence of acoustic features (e.g., LFCC) into a single score for one input trial, 
but the description is generalizable to those using a waveform input \cite{tak2020end, dinkel2017end}.
We use $\bs{x}_{1:N^{(j)}} \equiv (\bs{x}_1, \bs{x}_2, \cdots, \bs{x}_{N^{(j)}})\in\mathbb{R}^{N^{(j)}\times{D}}$ to denote the input feature sequence of the $j$-th trial, where $\bs{x}_n\in\mathbb{R}^{D}$ is the feature vector at the $n$-th frame, and where $N^{(j)}$ is the total number of frames.
A CM converts $\bs{x}_{1:N^{(j)}}$ into a score $s_j\in\mathbb{R}$ that shows how likely the input trial is bonafide. 

\subsection{Neural-network-based back end}
\label{seq:nn_structure}
\begin{figure}[t]
\centering
\includegraphics[trim=0 420 0 85, clip, width=0.5\textwidth]{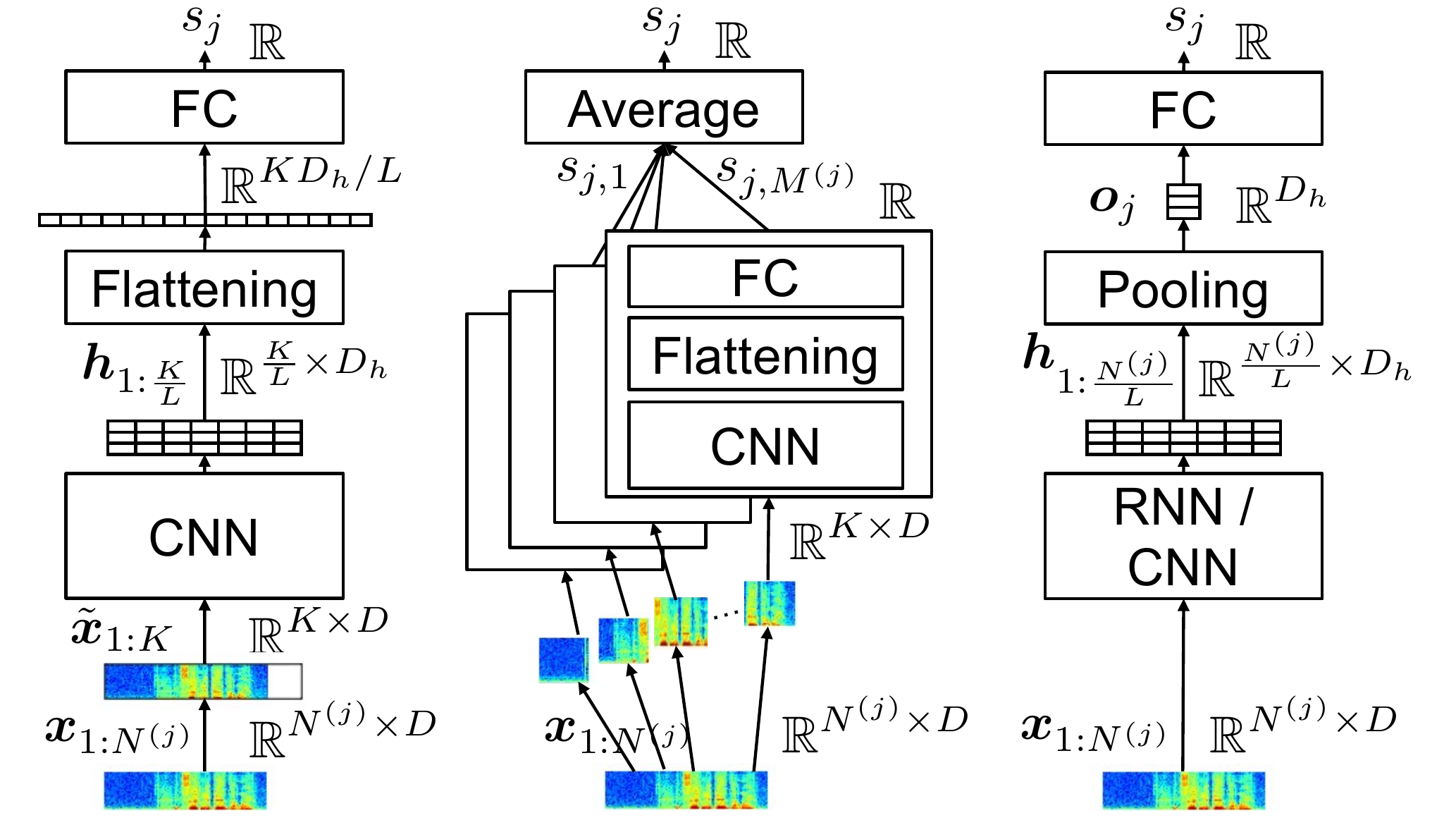}
\vspace{-5mm}
\caption{Typical neural CM backends during inference stage. FC denotes a fully-connected layer. Letter $j$ is a trial index.}
\label{fig:nn-struct}
\end{figure}

A practical issue is that $\bs{x}_{1:N^{(j)}}$ has varied length $N^{(j)}$. 
The strategy to handle the varied-length input depends on the back end, and Fig.~\ref{fig:nn-struct} illustrates three typical cases. 

\subsubsection{From fixed-length input to score}
The first strategy assumes a fixed-size input. It pads or trims $\bs{x}_{1:N^{(j)}}$ into $\tilde{\bs{x}}_{1:K}\in\mathbb{R}^{K\times{D}}$ and uses a CNN to transform $\tilde{\bs{x}}_{1:K}$ into $\bs{h}_{1:K/L}$, where $L$ is determined by convolution stride. It then flattens $\bs{h}_{1:K/L}$ and transforms it into the score $s_j$. 
The mapping can be summarized as $f: \mathbb{R}^{K\times{D}} \rightarrow \mathbb{R}^{{K/L}\times{D_h}} \rightarrow \mathbb{R}$.
This strategy is used in LCNN, ResNet, or other CNNs-based CMs \cite{lavrentyeva2019stc,zhang2020one,Chettri2019,lai2019attentive,Wu2020}.
The CM may pad short trials with random noises or replicated frames \cite{zhang2020one,lai2019attentive}. 
For trials with $N^{(j)} > K$, the CM may set $\tilde{\bs{x}}_{1:K} = \bs{x}_{n:n+K}$, where $n$ is a random integer \cite{zhang2020one} or is equal to one \cite{lavrentyeva2019stc}. 

Based on the above strategy,  the second approach usually sets a small $K$ and frames $\bs{x}_{1:N^{(j)}}$ into fixed-size chunks $(\bs{X}_1, \bs{X}_2, \cdots, \bs{X}_{M^{(j)}})$, where $\bs{X}_m \in\mathbb{R}^{K\times{D}}$. 
It then uses CNNs to convert each chunk $\bs{X}_{m}$ into a score $s_{j,m}$ and computes a final score $s_j= \frac{1}{M^{(j)}}\sum_{m} s_{j,m}$. Some CMs may also pad $\bs{x}_{1:N^{(j)}}$ into a long sequence before chunking \cite{Lai2019}. 

\subsubsection{From varied-length input to score}
Using a fixed-size input has side-effects: trimming discards information in $\bs{x}_{K+1:N^{(j)}}$, and padding propagates artifacts \cite{chettri2020dataset}. While chunking keeps all the information, the chunks are independently scored by the neural network.  

 For speech input with varied length, an alternative strategy is $f:\mathbb{R}^{N^{(j)}\times{D}} \rightarrow \mathbb{R}^{{N^{(j)}/L}\times{D_h}} \rightarrow \mathbb{R}^{D_h} \rightarrow \mathbb{R}$.
The first step to convert $\bs{x}_{1:N^{(j)}}$ to $\bs{h}_{1:N^{(j)}/L}$ is supported by common neural networks, where $L>1$ if it is a CNN with a large stride or $L=1$ for a recurrent neural network (RNN). 
The hidden $\bs{h}_{1:N^{(j)}/L}$  can be further pooled in an utterance-level vector $\bs{o}_j = \sum_{m=1}^{N^{(j)}/L}w_m\bs{h}_{m}$, and this $\bs{o}_j\in{\mathbb{R}^{D_h}}$ can be easily transformed into the score $s$. 
The pooling weight $w_m$  can be either uniform or computed by attention \cite{Zhu2018} \footnote{We may use RNN to process $\bs{h}_{1:N^{(j)}/L}$ and use the output from the last RNN step as $\bs{o}_k$ \cite{dinkel2017end, zhang2017investigation,Alluri2017}. However, this approach was found to be inferior to pooling-based ones \cite{dinkel2018investigating}.}. 
This strategy has been used in  a few RNN \cite{dinkel2018investigating} and CNN-based CMs \cite{Cai2019,Yang2019}.

\subsection{Loss functions}
\label{seq:loss_func}
\subsubsection{Cross entropy with vanilla and margin-based softmax}
\label{seq:loss_func_softmax}
The CM can be trained by minimizing cross entropy (CE). 
If the $j$-th trial has a class label $y_j\in\{1, \cdots, C\}$, the loss function over a dataset $\mathcal{D}$ is defined as
\begin{align}
\mathcal{L}^{(\text{ce})} &= -\frac{1}{|\mathcal{D}|}\sum_{j=1}^{|\mathcal{D}|}\sum_{k=1}^{C} \mathds{1}(y_j = k) \log P_{j, k},
\end{align}
where $P_{j, k}$ is the probability that the $j$-th trial is from class $k$ and can be computed through a softmax function. 
With the rightmost CM in Fig.~\ref{fig:nn-struct} as an example, we have $P_{j, k} = \frac{\exp(\boldsymbol{c}_{k}^{\top}\bs{o}_j)}{\sum_{i=1}^{C}\exp(\bs{c}_{i}^{\top}\boldsymbol{o}_j)}$,  where $\bs{c}_{k}\in\mathbb{R}^{D_h}$ is a weight vector for class $k$. 
When $C=2$, we can re-write $P_{j, k}$ as a sigmoid function, e.g., $P_{j,1} = \frac{1}{1 + \exp(-(\bs{c}_{1} - \bs{c}_{2})^{\top}\boldsymbol{o}_j))}$. This is widely used in CMs that assign either \textit{bonafide} or \textit{spoof} to the input trial. 

Advanced CMs enhance the softmax with margins, which can be written in general as \cite{deng2019arcface}:
\begin{align} 
P_{j, k} &= \frac{e^{\alpha[\cos (m_1 \theta_{j, k} + m_2) - m_3]}}{ e^{\alpha [\cos (m_1 \theta_{j, k} + m_2) - m_3]} + \sum_{i=1, i\neq k}^{C} e^{\alpha\cos\theta_{j, i}}}, 
\label{eq:margin_softmax}
\end{align}
where $\cos\theta_{j, k} = \widehat{\bs{c}}_k^\top\widehat{\bs{o}}_j$ is the cosine distance between length-normalized vector $\widehat{\bs{c}}_k = \bs{c}_k/||\bs{c}_k||$ and $\widehat{\bs{o}}_j = \bs{o}_j/||\bs{c}_j||$, and $(\alpha, m_1, m_2, m_3)$ is a hyper-parameter set that defines the type of the margin. For example, an angular-softmax \cite{liu2017sphereface} defined by $(m_1>1, m_2=0, m_3=0)$ is used in an LCNN-based CM \cite{lavrentyeva2019stc}, while an additive-margin (AM) softmax \cite{wang2018cosface,wang2018additive} defined by $(m_1=1, m_2=0, m_3>0)$ is used in recent CMs \cite{zhang2020one, Chen2020Odyssey}. 
Note that Eq.~(\ref{eq:margin_softmax}) may be further extended, for example, by defining a $m_{3,k}$ for each class $k$. This function is referred to as a one-class (OC) softmax and has been used for CMs \cite{zhang2020one}. 

The score in the inference stage is set by either $s=P_{j,1}$ or $s=\widehat{\bs{c}}_1^\top\widehat{\bs{o}}_j$ if class label $y_j=1$ denotes being bonafide. 

\begin{figure}[t]
\centering
\includegraphics[trim=0 540 0 85, clip, width=0.5\textwidth]{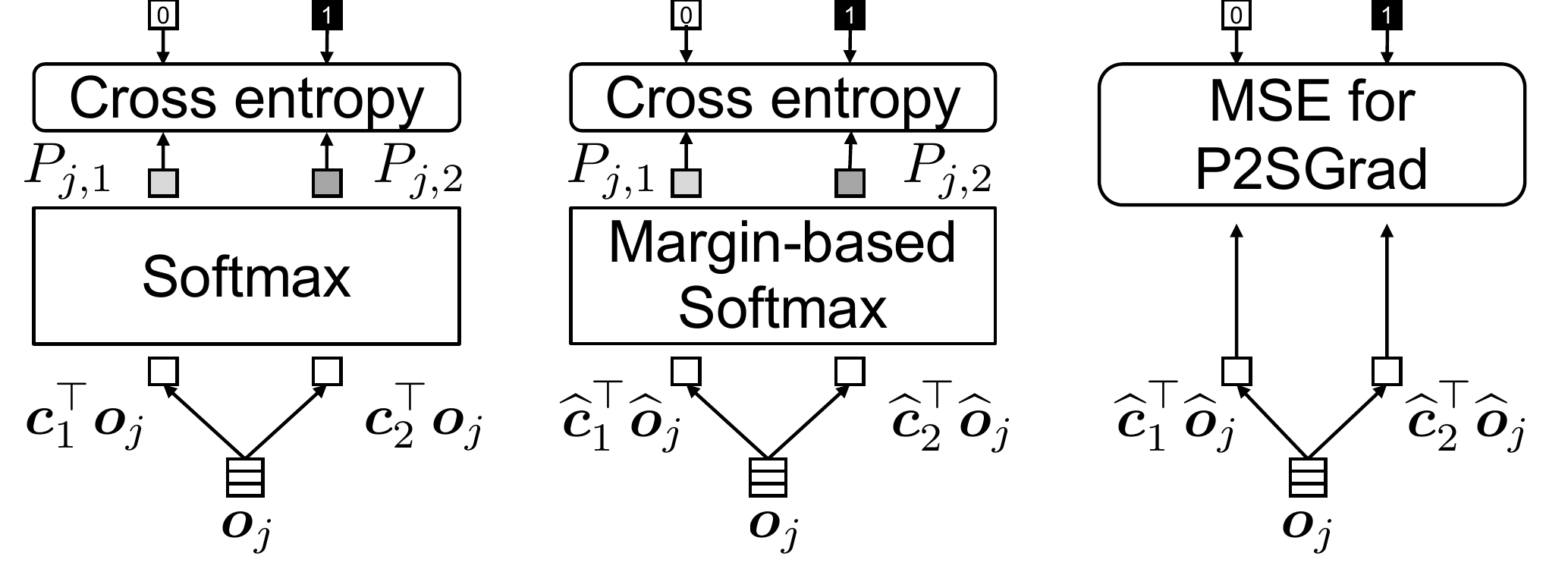}
\vspace{-5mm}
\caption{Loss functions for neural CM back ends and binary classification. Vector $o_j$ is from rightmost network in Fig.~\ref{fig:nn-struct}. }
\label{fig:nn-loss}
\end{figure}

\subsubsection{New mean-square-error loss function with P2SGrad}
\label{seq:loss_func_p2s}
The discriminative power of the margin-based softmax has been testified in a few speech \cite{Liu2019} 
and image processing tasks \cite{zhang2020one}, but its performance is sensitive to the hyper-parameter setting \cite{zhang2019p2sgrad}. 
Here, we propose a hyper-parameter-free loss function:
\begin{equation}
\mathcal{L}^{(\text{p2s})} = \frac{1}{|\mathcal{D}|}\sum_{j=1}^{|\mathcal{D}|}\sum_{k=1}^{C} (\cos\theta_{j,k} - \mathds{1}(y_j=k)) ^ 2,
\label{eq:p2sgrad}
\end{equation}
where $\cos\theta_{j, k} = \widehat{\bs{c}}_k^\top\widehat{\bs{o}}_j$. 
This loss is a mean-square error (MSE) between a network's output and a scalar target value, but the network's output here is a cosine distance $\cos\theta_{j, k} $ rather than an unconstrained value. The gradient ${\partial\mathcal{L}^{(\text{p2s})}}/{\partial\bs{o}_j}$ can be shown to be  identical to the probability-to-similarity gradient (P2SGrad) \cite{zhang2019p2sgrad}, and this points to the same optimal direction as the margin-based softmax. 
We refer to the new loss function as \textit{MSE for P2SGrad}.

Note that the P2SGrad paper defines only the gradient but not the loss function \cite{zhang2019p2sgrad}. 
During inference, we may set $s=\cos\theta_{j,1}$ as the output score if $k=1$ indicates the bonafide class.

\section{Experiments}
\label{seq:exp}

We conducted the experiments on the ASVspoof 2019 LA database \cite{WANG2020101114} and followed the official protocol without data augmentation. We released the code and all trained models under BSD-3-Clause License \footnote{https://github.com/nii-yamagishilab/project-NN-Pytorch-scripts}. Results that could not fit into this paper are included in the appendix on Arxiv\footnote{Arxiv link: https://arxiv.org/abs/2103.11326}.
%

\subsection{Experimental CMs}
\label{seq:exp_models}
Experimental CMs were from a combination of the network structures and loss functions described in Sec.~\ref{sec:overview}, but not all of them were included due to the limited computation resources: 
\begin{itemize}
\item \systemI: the same LCNN as \cite{lavrentyeva2019stc}, but the fixed-size input $\tilde{\bs{x}}_{1:K}$ used $K=750$ \cite{zhang2020one}. A shorter trial was padded with zero, and a longer trial was trimmed by $\tilde{\bs{x}}_{1:K} = \bs{x}_{n:n+K}$, where $n$ is a random integer;
\item \systemII: the same LCNN as \cite{lavrentyeva2019stc}, but layers after the CNN part\footnote{Specifically layers after `MaxPool\_28' in \cite{lavrentyeva2019stc}} were replaced with a single-head-attention pooling layer \cite{Zhu2018} and an FC layer;
\item \systemIII: the same LCNN as \cite{lavrentyeva2019stc}, but layers after the CNN part were replaced with two Bi-LSTM layers, an average pooling layer, and an FC layer. A skip connection was added over the two Bi-LSTM layers, and the size of the Bi-LSTM layers was equal to the dimensions of the CNN part's output. 
\end{itemize}
Both \systemII\ and \systemIII\ received varied-length input during training and inference, while \systemI\ processed only $\tilde{\bs{x}}_{1:K}$ for each trial.
Note that $K=750$ is long enough to cover the input features of 98\% of the trials in the database when using front ends explained later. 
Stacking attention to \systemIII\ was found to be unhelpful in a pilot experiment, while removing LSTMs from \systemIII\ degraded performance. These two configurations were thus not included in any further experiment.
The network in the middle of Fig.~\ref{fig:nn-struct} was excluded because of the cost of searching for good settings to frame the input feature. 

All networks were trained with CE for binary classification, and we assumed $y={1}$ and $y=2$ denote \textit{bonafide} and \textit{spoof}, respectively. The activation function can be either sigmoid, AM-softmax with $(\alpha=20, m_3=0.9)$ \cite{zhang2020one}, OC-softmax with $(\alpha=20, m_{3,1}=0.9, m_{3,2}=0.2)$ \cite{zhang2020one}, or MSE for P2SGrad. 
The latter three activation functions had an additional FC layer of size 64 in the front.
Accordingly, if we use the rightmost CM in Fig~\ref{fig:nn-struct} as example, the mapping from input to output score becomes $\mathbb{R}^{N^{(j)}\times{D}}\rightarrow\mathbb{R}^{N^{(j)}/L\times{D}}\rightarrow\mathbb{R}^{D_h}\rightarrow\mathbb{R}^{64}\rightarrow\mathbb{R}$.

Each network was tested with three front ends: LFCC, LFB, and spectrogram. 
The 60-dimensional LFCC followed the ASVspoof 2019 baseline recipe \cite{Todisco2019}: a frame length of 20ms,  a frame shift of 10ms, a 512-point FFT, a linearly spaced triangle filter bank of 20 channels, and delta plus delta-delta coefficients. However, the first dimension was replaced with log spectral energy.  The LFB had a similar configuration but contained only static coefficients from 60 linear filter-bank channels.
The spectrogram was configured similarly and had 257 dimensions. 

When using the spectrogram, a trainable FC layer initialized with coefficients of the linear filter bank for LFB was added before the LCNN. This layer compressed the spectrogram to 60-dimensional hidden features, and it improved the CMs performance by a large margin. Other detailed implementations that cannot be presented here can be found in the released Pytorch codes.

\subsection{Training recipe}
All the CMs were trained using an Adam optimizer with $\beta_1=0.9, \beta_2=0.999, \epsilon=10^{-8}$ \cite{kingma2014adam}. The initial learning rate of $3\times10^{-4}$ was multiplied by 0.5 for every ten epochs \cite{zhang2020one}. The mini-batch size was 64 or 8, and each mini-batch contained randomly selected trials with similar duration.  
No voice activity detection or feature normalization was used. 

It is known that CM performance varies when the training starts with a different random seed. To fairly compare the CMs with limited computation resources, we trained and evaluated each CM for six times with the random seed for the $k$-th run set to $10^{k-1}$. 
Note that no randomness occurred during the evaluation in our implementation except the random trimming in \systemI\ , but the randomness was trivial because only 2\% of the evaluation trials needed to be trimmed.
All the experiments were done using Nvidia Tesla P100. The results are reproducible with the same random seed and GPU environment.

\subsection{Evaluation metrics and statistical analysis}
We evaluated the CMs by calculating the EERs and tandem detection cost function (t-DCF) \cite{kinnunen2018t} using routines from the ASVspoof official website\footnote{https://www.asvspoof.org/asvspoof2019/tDCF\_python\_v1.zip}. 

We also conducted pair-wise statistical analysis using the methodology in \cite{bengio2004statistical}. For a pair of models $(A, B)$ in the comparison, a $z$ value was computed using 
\begin{equation}
z=\frac{2|\text{EER}_A - \text{EER}_B|}{\sqrt{\Big[\text{EER}_A (1-\text{EER}_A)+\text{EER}_B (1-\text{EER}_B)\Big]\frac{N_{\text{bona}} + N_{\text{spoof}}}{N_{\text{bona}} N_{\text{spoof}}}}},
\end{equation}
where $N_{\text{bona}}$ and $N_{\text{spoof}}$ denote the number of evaluated bonafide and spoof trials, respectively. 
$z$ was compared with a value $Z_{\alpha/2}$ decided by a significance level $\alpha=0.05$ and Holm-Bonferroni correction \cite{holm1979simple}, and $z\geq Z_{\alpha/2}$ suggests a statistically significant difference between $A$ and $B$.

\begin{table*}[t!]
\caption{EER and min t-DCF on ASVspoof2019 LA evaluation set. For visualization, the results of six training-evaluation rounds were sorted in accord with EER from low (I) to high (VI). A darker cell color indicates a higher EER or min t-DCF value.}
\vspace{-6mm}
\begin{center}
\resizebox{\textwidth}{!}{
\setlength{\tabcolsep}{2pt}
\begin{tabular}{cc|cccccc|cccccc|cccccc|cccccc}
\multicolumn{26}{c}{EERs}   \\
\toprule
  \multicolumn{2}{c}{} &  \multicolumn{6}{c}{AM-softmax} &  \multicolumn{6}{c}{OC-softmax} &  \multicolumn{6}{c}{Sigmoid} &  \multicolumn{6}{c}{MSE for P2SGrad}   \\
  \cmidrule(r){3-26}
\multicolumn{2}{c}{}   &  I  &  II  &  III   &  IV   &  V   &  \multicolumn{1}{c}{VI}   & I  &  II  &  III   &  IV   &  V   &   \multicolumn{1}{c}{VI}      & I  &  II  &  III   &  IV   &  V   &   \multicolumn{1}{c}{VI}       &  I  &  II  &  III   &  IV   &  V   &   \multicolumn{1}{c}{VI}   \\ 
\midrule

 \multirow{3}{*}{\rotatebox[origin=c]{0}{LFB}}  &  \systemI    & \cellcolor[rgb]{0.71, 0.71, 0.71} 5.59 & \cellcolor[rgb]{0.62, 0.62, 0.62} 6.39 & \cellcolor[rgb]{0.56, 0.56, 0.56} 7.12 & \cellcolor[rgb]{0.50, 0.50, 0.50} 7.79 & \cellcolor[rgb]{0.38, 0.38, 0.38} 9.33 & \cellcolor[rgb]{0.27, 0.27, 0.27} 10.72 & \cellcolor[rgb]{0.66, 0.66, 0.66} 5.98 & \cellcolor[rgb]{0.59, 0.59, 0.59} 6.76 & \cellcolor[rgb]{0.56, 0.56, 0.56} 7.02 & \cellcolor[rgb]{0.55, 0.55, 0.55} 7.14 & \cellcolor[rgb]{0.36, 0.36, 0.36} 9.56 & \cellcolor[rgb]{0.22, 0.22, 0.22} 11.34 & \cellcolor[rgb]{0.56, 0.56, 0.56} 7.00 & \cellcolor[rgb]{0.53, 0.53, 0.53} 7.40 & \cellcolor[rgb]{0.43, 0.43, 0.43} 8.69 & \cellcolor[rgb]{0.41, 0.41, 0.41} 8.89 & \cellcolor[rgb]{0.34, 0.34, 0.34} 9.86 & \cellcolor[rgb]{0.33, 0.33, 0.33} 10.13 & \cellcolor[rgb]{0.58, 0.58, 0.58} 6.81 & \cellcolor[rgb]{0.54, 0.54, 0.54} 7.25 & \cellcolor[rgb]{0.47, 0.47, 0.47} 8.10 & \cellcolor[rgb]{0.46, 0.46, 0.46} 8.28 & \cellcolor[rgb]{0.45, 0.45, 0.45} 8.41 & \cellcolor[rgb]{0.32, 0.32, 0.32} 10.21\\ 
                                                 &  \systemII   & \cellcolor[rgb]{0.82, 0.82, 0.82} 4.26 & \cellcolor[rgb]{0.82, 0.82, 0.82} 4.32 & \cellcolor[rgb]{0.79, 0.79, 0.79} 4.69 & \cellcolor[rgb]{0.74, 0.74, 0.74} 5.23 & \cellcolor[rgb]{0.44, 0.44, 0.44} 8.55 & \cellcolor[rgb]{0.01, 0.01, 0.01} \textcolor[rgb]{0.6,0.6,0.6}{14.86} & \cellcolor[rgb]{0.84, 0.84, 0.84} 4.01 & \cellcolor[rgb]{0.80, 0.80, 0.80} 4.54 & \cellcolor[rgb]{0.80, 0.80, 0.80} 4.57 & \cellcolor[rgb]{0.74, 0.74, 0.74} 5.27 & \cellcolor[rgb]{0.73, 0.73, 0.73} 5.33 & \cellcolor[rgb]{0.69, 0.69, 0.69} 5.76 & \cellcolor[rgb]{0.90, 0.90, 0.90} 3.34 & \cellcolor[rgb]{0.76, 0.76, 0.76} 5.07 & \cellcolor[rgb]{0.70, 0.70, 0.70} 5.70 & \cellcolor[rgb]{0.67, 0.67, 0.67} 5.91 & \cellcolor[rgb]{0.67, 0.67, 0.67} 5.93 & \cellcolor[rgb]{0.67, 0.67, 0.67} 5.93 & \cellcolor[rgb]{0.85, 0.85, 0.85} 3.99 & \cellcolor[rgb]{0.77, 0.77, 0.77} 4.87 & \cellcolor[rgb]{0.69, 0.69, 0.69} 5.75 & \cellcolor[rgb]{0.68, 0.68, 0.68} 5.85 & \cellcolor[rgb]{0.64, 0.64, 0.64} 6.20 & \cellcolor[rgb]{0.53, 0.53, 0.53} 7.36\\ 
                                                 &  \systemIII  & \cellcolor[rgb]{0.83, 0.83, 0.83} 4.24 & \cellcolor[rgb]{0.74, 0.74, 0.74} 5.27 & \cellcolor[rgb]{0.69, 0.69, 0.69} 5.71 & \cellcolor[rgb]{0.64, 0.64, 0.64} 6.23 & \cellcolor[rgb]{0.56, 0.56, 0.56} 7.03 & \cellcolor[rgb]{0.56, 0.56, 0.56} 7.10 & \cellcolor[rgb]{0.68, 0.68, 0.68} 5.81 & \cellcolor[rgb]{0.61, 0.61, 0.61} 6.51 & \cellcolor[rgb]{0.57, 0.57, 0.57} 6.89 & \cellcolor[rgb]{0.51, 0.51, 0.51} 7.64 & \cellcolor[rgb]{0.39, 0.39, 0.39} 9.15 & \cellcolor[rgb]{0.32, 0.32, 0.32} 10.24 & \cellcolor[rgb]{0.56, 0.56, 0.56} 7.04 & \cellcolor[rgb]{0.49, 0.49, 0.49} 7.93 & \cellcolor[rgb]{0.44, 0.44, 0.44} 8.52 & \cellcolor[rgb]{0.39, 0.39, 0.39} 9.16 & \cellcolor[rgb]{0.37, 0.37, 0.37} 9.53 & \cellcolor[rgb]{0.34, 0.34, 0.34} 9.84 & \cellcolor[rgb]{0.76, 0.76, 0.76} 5.06 & \cellcolor[rgb]{0.75, 0.75, 0.75} 5.17 & \cellcolor[rgb]{0.70, 0.70, 0.70} 5.69 & \cellcolor[rgb]{0.69, 0.69, 0.69} 5.75 & \cellcolor[rgb]{0.64, 0.64, 0.64} 6.17 & \cellcolor[rgb]{0.61, 0.61, 0.61} 6.50\\ 
\midrule
 \multirow{3}{*}{\rotatebox[origin=c]{0}{SPEC}} &  \systemI    & \cellcolor[rgb]{0.77, 0.77, 0.77} 4.84 & \cellcolor[rgb]{0.76, 0.76, 0.76} 5.02 & \cellcolor[rgb]{0.74, 0.74, 0.74} 5.23 & \cellcolor[rgb]{0.68, 0.68, 0.68} 5.87 & \cellcolor[rgb]{0.67, 0.67, 0.67} 5.90 & \cellcolor[rgb]{0.64, 0.64, 0.64} 6.25 & \cellcolor[rgb]{0.81, 0.81, 0.81} 4.41 & \cellcolor[rgb]{0.80, 0.80, 0.80} 4.60 & \cellcolor[rgb]{0.77, 0.77, 0.77} 4.84 & \cellcolor[rgb]{0.76, 0.76, 0.76} 5.04 & \cellcolor[rgb]{0.73, 0.73, 0.73} 5.37 & \cellcolor[rgb]{0.63, 0.63, 0.63} 6.35 & \cellcolor[rgb]{0.92, 0.92, 0.92} 3.09 & \cellcolor[rgb]{0.88, 0.88, 0.88} 3.58 & \cellcolor[rgb]{0.78, 0.78, 0.78} 4.72 & \cellcolor[rgb]{0.77, 0.77, 0.77} 4.82 & \cellcolor[rgb]{0.74, 0.74, 0.74} 5.22 & \cellcolor[rgb]{0.71, 0.71, 0.71} 5.56 & \cellcolor[rgb]{0.93, 0.93, 0.93} 2.94 & \cellcolor[rgb]{0.90, 0.90, 0.90} 3.22 & \cellcolor[rgb]{0.88, 0.88, 0.88} 3.59 & \cellcolor[rgb]{0.87, 0.87, 0.87} 3.73 & \cellcolor[rgb]{0.81, 0.81, 0.81} 4.49 & \cellcolor[rgb]{0.78, 0.78, 0.78} 4.74\\ 
                                                 &  \systemII   & \cellcolor[rgb]{0.84, 0.84, 0.84} 4.02 & \cellcolor[rgb]{0.84, 0.84, 0.84} 4.08 & \cellcolor[rgb]{0.76, 0.76, 0.76} 4.99 & \cellcolor[rgb]{0.74, 0.74, 0.74} 5.22 & \cellcolor[rgb]{0.71, 0.71, 0.71} 5.57 & \cellcolor[rgb]{0.47, 0.47, 0.47} 8.20 & \cellcolor[rgb]{0.84, 0.84, 0.84} 4.05 & \cellcolor[rgb]{0.80, 0.80, 0.80} 4.55 & \cellcolor[rgb]{0.77, 0.77, 0.77} 4.94 & \cellcolor[rgb]{0.70, 0.70, 0.70} 5.67 & \cellcolor[rgb]{0.63, 0.63, 0.63} 6.31 & \cellcolor[rgb]{0.44, 0.44, 0.44} 8.47 & \cellcolor[rgb]{0.85, 0.85, 0.85} 3.92 & \cellcolor[rgb]{0.84, 0.84, 0.84} 4.04 & \cellcolor[rgb]{0.81, 0.81, 0.81} 4.42 & \cellcolor[rgb]{0.77, 0.77, 0.77} 4.91 & \cellcolor[rgb]{0.77, 0.77, 0.77} 4.95 & \cellcolor[rgb]{0.60, 0.60, 0.60} 6.62 & \cellcolor[rgb]{0.78, 0.78, 0.78} 4.70 & \cellcolor[rgb]{0.76, 0.76, 0.76} 5.03 & \cellcolor[rgb]{0.71, 0.71, 0.71} 5.60 & \cellcolor[rgb]{0.68, 0.68, 0.68} 5.88 & \cellcolor[rgb]{0.63, 0.63, 0.63} 6.35 & \cellcolor[rgb]{0.61, 0.61, 0.61} 6.50\\ 
                                                 &  \systemIII  & \cellcolor[rgb]{0.85, 0.85, 0.85} 3.96 & \cellcolor[rgb]{0.84, 0.84, 0.84} 4.04 & \cellcolor[rgb]{0.81, 0.81, 0.81} 4.38 & \cellcolor[rgb]{0.80, 0.80, 0.80} 4.52 & \cellcolor[rgb]{0.75, 0.75, 0.75} 5.13 & \cellcolor[rgb]{0.67, 0.67, 0.67} 5.97 & \cellcolor[rgb]{0.94, 0.94, 0.94} 2.81 & \cellcolor[rgb]{0.89, 0.89, 0.89} 3.41 & \cellcolor[rgb]{0.80, 0.80, 0.80} 4.49 & \cellcolor[rgb]{0.80, 0.80, 0.80} 4.50 & \cellcolor[rgb]{0.79, 0.79, 0.79} 4.65 & \cellcolor[rgb]{0.77, 0.77, 0.77} 4.91 & \cellcolor[rgb]{0.90, 0.90, 0.90} 3.29 & \cellcolor[rgb]{0.88, 0.88, 0.88} 3.56 & \cellcolor[rgb]{0.86, 0.86, 0.86} 3.82 & \cellcolor[rgb]{0.81, 0.81, 0.81} 4.45 & \cellcolor[rgb]{0.80, 0.80, 0.80} 4.61 & \cellcolor[rgb]{0.72, 0.72, 0.72} 5.44 & \cellcolor[rgb]{0.97, 0.97, 0.97} 2.37 & \cellcolor[rgb]{0.93, 0.93, 0.93} 2.91 & \cellcolor[rgb]{0.92, 0.92, 0.92} 3.00 & \cellcolor[rgb]{0.85, 0.85, 0.85} 3.94 & \cellcolor[rgb]{0.82, 0.82, 0.82} 4.26 & \cellcolor[rgb]{0.81, 0.81, 0.81} 4.37\\ 
\midrule 
 \multirow{3}{*}{\rotatebox[origin=c]{0}{LFCC}} &  \systemI    & \cellcolor[rgb]{0.92, 0.92, 0.92} 3.04 & \cellcolor[rgb]{0.88, 0.88, 0.88} 3.52 & \cellcolor[rgb]{0.78, 0.78, 0.78} 4.73 & \cellcolor[rgb]{0.77, 0.77, 0.77} 4.89 & \cellcolor[rgb]{0.68, 0.68, 0.68} 5.85 & \cellcolor[rgb]{0.56, 0.56, 0.56} 7.06 & \cellcolor[rgb]{0.93, 0.93, 0.93} 2.93 & \cellcolor[rgb]{0.93, 0.93, 0.93} 2.93 & \cellcolor[rgb]{0.93, 0.93, 0.93} 2.95 & \cellcolor[rgb]{0.92, 0.92, 0.92} 2.99 & \cellcolor[rgb]{0.86, 0.86, 0.86} 3.84 & \cellcolor[rgb]{0.85, 0.85, 0.85} 4.00 & \cellcolor[rgb]{0.96, 0.96, 0.96} 2.54 & \cellcolor[rgb]{0.94, 0.94, 0.94} 2.73 & \cellcolor[rgb]{0.94, 0.94, 0.94} 2.77 & \cellcolor[rgb]{0.93, 0.93, 0.93} 2.91 & \cellcolor[rgb]{0.92, 0.92, 0.92} 3.08 & \cellcolor[rgb]{0.89, 0.89, 0.89} 3.47 & \cellcolor[rgb]{0.97, 0.97, 0.97} 2.31 & \cellcolor[rgb]{0.96, 0.96, 0.96} 2.46 & \cellcolor[rgb]{0.95, 0.95, 0.95} 2.64 & \cellcolor[rgb]{0.95, 0.95, 0.95} 2.65 & \cellcolor[rgb]{0.92, 0.92, 0.92} 3.09 & \cellcolor[rgb]{0.91, 0.91, 0.91} 3.11\\ 
                                                 &  \systemII   & \cellcolor[rgb]{0.92, 0.92, 0.92} 2.99 & \cellcolor[rgb]{0.88, 0.88, 0.88} 3.48 & \cellcolor[rgb]{0.88, 0.88, 0.88} 3.60 & \cellcolor[rgb]{0.88, 0.88, 0.88} 3.61 & \cellcolor[rgb]{0.88, 0.88, 0.88} 3.62 & \cellcolor[rgb]{0.85, 0.85, 0.85} 3.92 & \cellcolor[rgb]{0.93, 0.93, 0.93} 2.91 & \cellcolor[rgb]{0.93, 0.93, 0.93} 2.96 & \cellcolor[rgb]{0.89, 0.89, 0.89} 3.36 & \cellcolor[rgb]{0.89, 0.89, 0.89} 3.40 & \cellcolor[rgb]{0.88, 0.88, 0.88} 3.61 & \cellcolor[rgb]{0.87, 0.87, 0.87} 3.71 & \cellcolor[rgb]{0.91, 0.91, 0.91} 3.18 & \cellcolor[rgb]{0.91, 0.91, 0.91} 3.19 & \cellcolor[rgb]{0.90, 0.90, 0.90} 3.24 & \cellcolor[rgb]{0.90, 0.90, 0.90} 3.29 & \cellcolor[rgb]{0.89, 0.89, 0.89} 3.39 & \cellcolor[rgb]{0.84, 0.84, 0.84} 4.12 & \cellcolor[rgb]{0.94, 0.94, 0.94} 2.72 & \cellcolor[rgb]{0.94, 0.94, 0.94} 2.81 & \cellcolor[rgb]{0.93, 0.93, 0.93} 2.91 & \cellcolor[rgb]{0.92, 0.92, 0.92} 3.03 & \cellcolor[rgb]{0.90, 0.90, 0.90} 3.22 & \cellcolor[rgb]{0.90, 0.90, 0.90} 3.30\\ 
                                                 &  \systemIII  & \cellcolor[rgb]{0.96, 0.96, 0.96} 2.46 & \cellcolor[rgb]{0.92, 0.92, 0.92} 3.02 & \cellcolor[rgb]{0.90, 0.90, 0.90} 3.23 & \cellcolor[rgb]{0.90, 0.90, 0.90} 3.34 & \cellcolor[rgb]{0.89, 0.89, 0.89} 3.41 & \cellcolor[rgb]{0.86, 0.86, 0.86} 3.86 & \cellcolor[rgb]{0.97, 0.97, 0.97} 2.23 & \cellcolor[rgb]{0.96, 0.96, 0.96} 2.42 & \cellcolor[rgb]{0.93, 0.93, 0.93} 2.96 & \cellcolor[rgb]{0.93, 0.93, 0.93} 2.96 & \cellcolor[rgb]{0.93, 0.93, 0.93} 2.96 & \cellcolor[rgb]{0.79, 0.79, 0.79} 4.64 & \cellcolor[rgb]{0.95, 0.95, 0.95} 2.67 & \cellcolor[rgb]{0.93, 0.93, 0.93} 2.94 & \cellcolor[rgb]{0.91, 0.91, 0.91} 3.20 & \cellcolor[rgb]{0.89, 0.89, 0.89} 3.40 & \cellcolor[rgb]{0.89, 0.89, 0.89} 3.45 & \cellcolor[rgb]{0.88, 0.88, 0.88} 3.53 & \cellcolor[rgb]{1.00, 1.00, 1.00} 1.92 & \cellcolor[rgb]{0.99, 0.99, 0.99} 2.09 & \cellcolor[rgb]{0.96, 0.96, 0.96} 2.43 & \cellcolor[rgb]{0.96, 0.96, 0.96} 2.50 & \cellcolor[rgb]{0.95, 0.95, 0.95} 2.62 & \cellcolor[rgb]{0.92, 0.92, 0.92} 3.10\\ 
 \bottomrule
 \\
\multicolumn{26}{c}{min t-DCF (legacy version used in ASVspoof 2019 challenge)}   \\
\toprule
 \multirow{3}{*}{\rotatebox[origin=c]{0}{LFB}}  &  \systemI         & \cellcolor[rgb]{0.72, 0.72, 0.72} 0.129 & \cellcolor[rgb]{0.54, 0.54, 0.54} 0.168 & \cellcolor[rgb]{0.41, 0.41, 0.41} 0.204 & \cellcolor[rgb]{0.38, 0.38, 0.38} 0.212 & \cellcolor[rgb]{0.30, 0.30, 0.30} 0.235 & \cellcolor[rgb]{0.30, 0.30, 0.30} 0.236 & \cellcolor[rgb]{0.59, 0.59, 0.59} 0.156 & \cellcolor[rgb]{0.50, 0.50, 0.50} 0.179 & \cellcolor[rgb]{0.52, 0.52, 0.52} 0.173 & \cellcolor[rgb]{0.53, 0.53, 0.53} 0.171 & \cellcolor[rgb]{0.42, 0.42, 0.42} 0.200 & \cellcolor[rgb]{0.26, 0.26, 0.26} 0.246 & \cellcolor[rgb]{0.55, 0.55, 0.55} 0.166 & \cellcolor[rgb]{0.51, 0.51, 0.51} 0.175 & \cellcolor[rgb]{0.44, 0.44, 0.44} 0.195 & \cellcolor[rgb]{0.35, 0.35, 0.35} 0.222 & \cellcolor[rgb]{0.32, 0.32, 0.32} 0.229 & \cellcolor[rgb]{0.33, 0.33, 0.33} 0.227 & \cellcolor[rgb]{0.53, 0.53, 0.53} 0.171 & \cellcolor[rgb]{0.38, 0.38, 0.38} 0.212 & \cellcolor[rgb]{0.42, 0.42, 0.42} 0.200 & \cellcolor[rgb]{0.46, 0.46, 0.46} 0.190 & \cellcolor[rgb]{0.36, 0.36, 0.36} 0.217 & \cellcolor[rgb]{0.37, 0.37, 0.37} 0.215\\ 
                                                &  \systemII        & \cellcolor[rgb]{0.76, 0.76, 0.76} 0.118 & \cellcolor[rgb]{0.77, 0.77, 0.77} 0.117 & \cellcolor[rgb]{0.86, 0.86, 0.86} 0.093 & \cellcolor[rgb]{0.83, 0.83, 0.83} 0.101 & \cellcolor[rgb]{0.41, 0.41, 0.41} 0.203 & \cellcolor[rgb]{0.01, 0.01, 0.01} \textcolor[rgb]{0.6,0.6,0.6}{0.331} & \cellcolor[rgb]{0.79, 0.79, 0.79} 0.111 & \cellcolor[rgb]{0.75, 0.75, 0.75} 0.121 & \cellcolor[rgb]{0.75, 0.75, 0.75} 0.122 & \cellcolor[rgb]{0.62, 0.62, 0.62} 0.148 & \cellcolor[rgb]{0.59, 0.59, 0.59} 0.155 & \cellcolor[rgb]{0.57, 0.57, 0.57} 0.160 & \cellcolor[rgb]{0.96, 0.96, 0.96} 0.064 & \cellcolor[rgb]{0.88, 0.88, 0.88} 0.087 & \cellcolor[rgb]{0.73, 0.73, 0.73} 0.126 & \cellcolor[rgb]{0.88, 0.88, 0.88} 0.088 & \cellcolor[rgb]{0.80, 0.80, 0.80} 0.108 & \cellcolor[rgb]{0.86, 0.86, 0.86} 0.092 & \cellcolor[rgb]{0.87, 0.87, 0.87} 0.089 & \cellcolor[rgb]{0.80, 0.80, 0.80} 0.110 & \cellcolor[rgb]{0.74, 0.74, 0.74} 0.124 & \cellcolor[rgb]{0.76, 0.76, 0.76} 0.119 & \cellcolor[rgb]{0.74, 0.74, 0.74} 0.123 & \cellcolor[rgb]{0.71, 0.71, 0.71} 0.131\\ 
                                                &  \systemIII       & \cellcolor[rgb]{0.87, 0.87, 0.87} 0.091 & \cellcolor[rgb]{0.66, 0.66, 0.66} 0.141 & \cellcolor[rgb]{0.63, 0.63, 0.63} 0.146 & \cellcolor[rgb]{0.43, 0.43, 0.43} 0.197 & \cellcolor[rgb]{0.44, 0.44, 0.44} 0.195 & \cellcolor[rgb]{0.43, 0.43, 0.43} 0.197 & \cellcolor[rgb]{0.57, 0.57, 0.57} 0.160 & \cellcolor[rgb]{0.48, 0.48, 0.48} 0.184 & \cellcolor[rgb]{0.55, 0.55, 0.55} 0.165 & \cellcolor[rgb]{0.43, 0.43, 0.43} 0.198 & \cellcolor[rgb]{0.39, 0.39, 0.39} 0.210 & \cellcolor[rgb]{0.11, 0.11, 0.11} \textcolor[rgb]{0.6,0.6,0.6}{0.285} & \cellcolor[rgb]{0.49, 0.49, 0.49} 0.182 & \cellcolor[rgb]{0.42, 0.42, 0.42} 0.201 & \cellcolor[rgb]{0.43, 0.43, 0.43} 0.197 & \cellcolor[rgb]{0.43, 0.43, 0.43} 0.197 & \cellcolor[rgb]{0.32, 0.32, 0.32} 0.231 & \cellcolor[rgb]{0.34, 0.34, 0.34} 0.224 & \cellcolor[rgb]{0.63, 0.63, 0.63} 0.148 & \cellcolor[rgb]{0.63, 0.63, 0.63} 0.147 & \cellcolor[rgb]{0.59, 0.59, 0.59} 0.156 & \cellcolor[rgb]{0.56, 0.56, 0.56} 0.163 & \cellcolor[rgb]{0.53, 0.53, 0.53} 0.172 & \cellcolor[rgb]{0.48, 0.48, 0.48} 0.183\\ 
\midrule
 \multirow{3}{*}{\rotatebox[origin=c]{0}{SPEC}} &  \systemI         & \cellcolor[rgb]{0.72, 0.72, 0.72} 0.129 & \cellcolor[rgb]{0.71, 0.71, 0.71} 0.132 & \cellcolor[rgb]{0.71, 0.71, 0.71} 0.130 & \cellcolor[rgb]{0.70, 0.70, 0.70} 0.133 & \cellcolor[rgb]{0.66, 0.66, 0.66} 0.141 & \cellcolor[rgb]{0.58, 0.58, 0.58} 0.158 & \cellcolor[rgb]{0.75, 0.75, 0.75} 0.123 & \cellcolor[rgb]{0.75, 0.75, 0.75} 0.122 & \cellcolor[rgb]{0.75, 0.75, 0.75} 0.122 & \cellcolor[rgb]{0.71, 0.71, 0.71} 0.130 & \cellcolor[rgb]{0.70, 0.70, 0.70} 0.134 & \cellcolor[rgb]{0.58, 0.58, 0.58} 0.158 & \cellcolor[rgb]{0.89, 0.89, 0.89} 0.085 & \cellcolor[rgb]{0.84, 0.84, 0.84} 0.099 & \cellcolor[rgb]{0.75, 0.75, 0.75} 0.123 & \cellcolor[rgb]{0.71, 0.71, 0.71} 0.132 & \cellcolor[rgb]{0.68, 0.68, 0.68} 0.137 & \cellcolor[rgb]{0.62, 0.62, 0.62} 0.148 & \cellcolor[rgb]{0.89, 0.89, 0.89} 0.085 & \cellcolor[rgb]{0.88, 0.88, 0.88} 0.088 & \cellcolor[rgb]{0.85, 0.85, 0.85} 0.095 & \cellcolor[rgb]{0.81, 0.81, 0.81} 0.106 & \cellcolor[rgb]{0.73, 0.73, 0.73} 0.126 & \cellcolor[rgb]{0.73, 0.73, 0.73} 0.127\\ 
                                                &  \systemII        & \cellcolor[rgb]{0.76, 0.76, 0.76} 0.119 & \cellcolor[rgb]{0.77, 0.77, 0.77} 0.117 & \cellcolor[rgb]{0.71, 0.71, 0.71} 0.131 & \cellcolor[rgb]{0.60, 0.60, 0.60} 0.153 & \cellcolor[rgb]{0.67, 0.67, 0.67} 0.139 & \cellcolor[rgb]{0.59, 0.59, 0.59} 0.156 & \cellcolor[rgb]{0.82, 0.82, 0.82} 0.105 & \cellcolor[rgb]{0.73, 0.73, 0.73} 0.126 & \cellcolor[rgb]{0.66, 0.66, 0.66} 0.141 & \cellcolor[rgb]{0.60, 0.60, 0.60} 0.153 & \cellcolor[rgb]{0.56, 0.56, 0.56} 0.162 & \cellcolor[rgb]{0.44, 0.44, 0.44} 0.194 & \cellcolor[rgb]{0.80, 0.80, 0.80} 0.109 & \cellcolor[rgb]{0.80, 0.80, 0.80} 0.109 & \cellcolor[rgb]{0.77, 0.77, 0.77} 0.115 & \cellcolor[rgb]{0.67, 0.67, 0.67} 0.140 & \cellcolor[rgb]{0.66, 0.66, 0.66} 0.141 & \cellcolor[rgb]{0.52, 0.52, 0.52} 0.173 & \cellcolor[rgb]{0.69, 0.69, 0.69} 0.135 & \cellcolor[rgb]{0.74, 0.74, 0.74} 0.125 & \cellcolor[rgb]{0.65, 0.65, 0.65} 0.143 & \cellcolor[rgb]{0.72, 0.72, 0.72} 0.129 & \cellcolor[rgb]{0.61, 0.61, 0.61} 0.153 & \cellcolor[rgb]{0.48, 0.48, 0.48} 0.183\\ 
                                                &  \systemIII       & \cellcolor[rgb]{0.81, 0.81, 0.81} 0.106 & \cellcolor[rgb]{0.83, 0.83, 0.83} 0.102 & \cellcolor[rgb]{0.77, 0.77, 0.77} 0.116 & \cellcolor[rgb]{0.76, 0.76, 0.76} 0.118 & \cellcolor[rgb]{0.69, 0.69, 0.69} 0.136 & \cellcolor[rgb]{0.70, 0.70, 0.70} 0.134 & \cellcolor[rgb]{0.92, 0.92, 0.92} 0.077 & \cellcolor[rgb]{0.88, 0.88, 0.88} 0.088 & \cellcolor[rgb]{0.76, 0.76, 0.76} 0.118 & \cellcolor[rgb]{0.72, 0.72, 0.72} 0.130 & \cellcolor[rgb]{0.75, 0.75, 0.75} 0.121 & \cellcolor[rgb]{0.73, 0.73, 0.73} 0.126 & \cellcolor[rgb]{0.88, 0.88, 0.88} 0.087 & \cellcolor[rgb]{0.81, 0.81, 0.81} 0.105 & \cellcolor[rgb]{0.83, 0.83, 0.83} 0.102 & \cellcolor[rgb]{0.75, 0.75, 0.75} 0.122 & \cellcolor[rgb]{0.75, 0.75, 0.75} 0.122 & \cellcolor[rgb]{0.63, 0.63, 0.63} 0.146 & \cellcolor[rgb]{0.97, 0.97, 0.97} 0.060 & \cellcolor[rgb]{0.92, 0.92, 0.92} 0.077 & \cellcolor[rgb]{0.91, 0.91, 0.91} 0.079 & \cellcolor[rgb]{0.85, 0.85, 0.85} 0.097 & \cellcolor[rgb]{0.84, 0.84, 0.84} 0.097 & \cellcolor[rgb]{0.77, 0.77, 0.77} 0.117\\ 
\midrule
 \multirow{3}{*}{\rotatebox[origin=c]{0}{LFCC}} &  \systemI         & \cellcolor[rgb]{0.95, 0.95, 0.95} 0.068 & \cellcolor[rgb]{0.90, 0.90, 0.90} 0.081 & \cellcolor[rgb]{0.90, 0.90, 0.90} 0.083 & \cellcolor[rgb]{0.86, 0.86, 0.86} 0.092 & \cellcolor[rgb]{0.87, 0.87, 0.87} 0.090 & \cellcolor[rgb]{0.76, 0.76, 0.76} 0.118 & \cellcolor[rgb]{0.95, 0.95, 0.95} 0.068 & \cellcolor[rgb]{0.95, 0.95, 0.95} 0.068 & \cellcolor[rgb]{0.94, 0.94, 0.94} 0.071 & \cellcolor[rgb]{0.88, 0.88, 0.88} 0.087 & \cellcolor[rgb]{0.82, 0.82, 0.82} 0.103 & \cellcolor[rgb]{0.84, 0.84, 0.84} 0.099 & \cellcolor[rgb]{0.95, 0.95, 0.95} 0.069 & \cellcolor[rgb]{0.95, 0.95, 0.95} 0.069 & \cellcolor[rgb]{0.94, 0.94, 0.94} 0.071 & \cellcolor[rgb]{0.93, 0.93, 0.93} 0.074 & \cellcolor[rgb]{0.93, 0.93, 0.93} 0.074 & \cellcolor[rgb]{0.84, 0.84, 0.84} 0.099 & \cellcolor[rgb]{0.98, 0.98, 0.98} 0.056 & \cellcolor[rgb]{0.97, 0.97, 0.97} 0.061 & \cellcolor[rgb]{0.95, 0.95, 0.95} 0.068 & \cellcolor[rgb]{0.96, 0.96, 0.96} 0.065 & \cellcolor[rgb]{0.93, 0.93, 0.93} 0.074 & \cellcolor[rgb]{0.93, 0.93, 0.93} 0.075\\ 
                                                &  \systemII        & \cellcolor[rgb]{0.93, 0.93, 0.93} 0.074 & \cellcolor[rgb]{0.93, 0.93, 0.93} 0.073 & \cellcolor[rgb]{0.94, 0.94, 0.94} 0.071 & \cellcolor[rgb]{0.92, 0.92, 0.92} 0.077 & \cellcolor[rgb]{0.94, 0.94, 0.94} 0.072 & \cellcolor[rgb]{0.93, 0.93, 0.93} 0.074 & \cellcolor[rgb]{0.96, 0.96, 0.96} 0.066 & \cellcolor[rgb]{0.93, 0.93, 0.93} 0.074 & \cellcolor[rgb]{0.93, 0.93, 0.93} 0.074 & \cellcolor[rgb]{0.92, 0.92, 0.92} 0.076 & \cellcolor[rgb]{0.89, 0.89, 0.89} 0.084 & \cellcolor[rgb]{0.89, 0.89, 0.89} 0.084 & \cellcolor[rgb]{0.95, 0.95, 0.95} 0.068 & \cellcolor[rgb]{0.90, 0.90, 0.90} 0.081 & \cellcolor[rgb]{0.92, 0.92, 0.92} 0.076 & \cellcolor[rgb]{0.92, 0.92, 0.92} 0.076 & \cellcolor[rgb]{0.88, 0.88, 0.88} 0.086 & \cellcolor[rgb]{0.83, 0.83, 0.83} 0.101 & \cellcolor[rgb]{0.91, 0.91, 0.91} 0.079 & \cellcolor[rgb]{0.91, 0.91, 0.91} 0.080 & \cellcolor[rgb]{0.91, 0.91, 0.91} 0.080 & \cellcolor[rgb]{0.91, 0.91, 0.91} 0.079 & \cellcolor[rgb]{0.90, 0.90, 0.90} 0.082 & \cellcolor[rgb]{0.87, 0.87, 0.87} 0.091\\ 
                                                &  \systemIII       & \cellcolor[rgb]{0.98, 0.98, 0.98} 0.057 & \cellcolor[rgb]{0.95, 0.95, 0.95} 0.068 & \cellcolor[rgb]{0.91, 0.91, 0.91} 0.078 & \cellcolor[rgb]{0.92, 0.92, 0.92} 0.077 & \cellcolor[rgb]{0.91, 0.91, 0.91} 0.079 & \cellcolor[rgb]{0.95, 0.95, 0.95} 0.068 & \cellcolor[rgb]{0.96, 0.96, 0.96} 0.064 & \cellcolor[rgb]{0.96, 0.96, 0.96} 0.064 & \cellcolor[rgb]{0.94, 0.94, 0.94} 0.073 & \cellcolor[rgb]{0.92, 0.92, 0.92} 0.076 & \cellcolor[rgb]{0.92, 0.92, 0.92} 0.076 & \cellcolor[rgb]{0.73, 0.73, 0.73} 0.127 & \cellcolor[rgb]{0.96, 0.96, 0.96} 0.064 & \cellcolor[rgb]{0.96, 0.96, 0.96} 0.063 & \cellcolor[rgb]{0.95, 0.95, 0.95} 0.066 & \cellcolor[rgb]{0.90, 0.90, 0.90} 0.083 & \cellcolor[rgb]{0.94, 0.94, 0.94} 0.073 & \cellcolor[rgb]{0.92, 0.92, 0.92} 0.075 & \cellcolor[rgb]{1.00, 1.00, 1.00} 0.052 & \cellcolor[rgb]{0.98, 0.98, 0.98} 0.058 & \cellcolor[rgb]{0.95, 0.95, 0.95} 0.067 & \cellcolor[rgb]{0.99, 0.99, 0.99} 0.055 & \cellcolor[rgb]{0.95, 0.95, 0.95} 0.067 & \cellcolor[rgb]{0.93, 0.93, 0.93} 0.075\\
  \bottomrule
\end{tabular}
}
\end{center}
\vspace{-6mm}
\label{tab:eer}
\end{table*}%

\subsection{Experimental results}
\begin{figure}[t]
\centering
\includegraphics[trim=0 170 0 180, clip, width=0.48\textwidth]{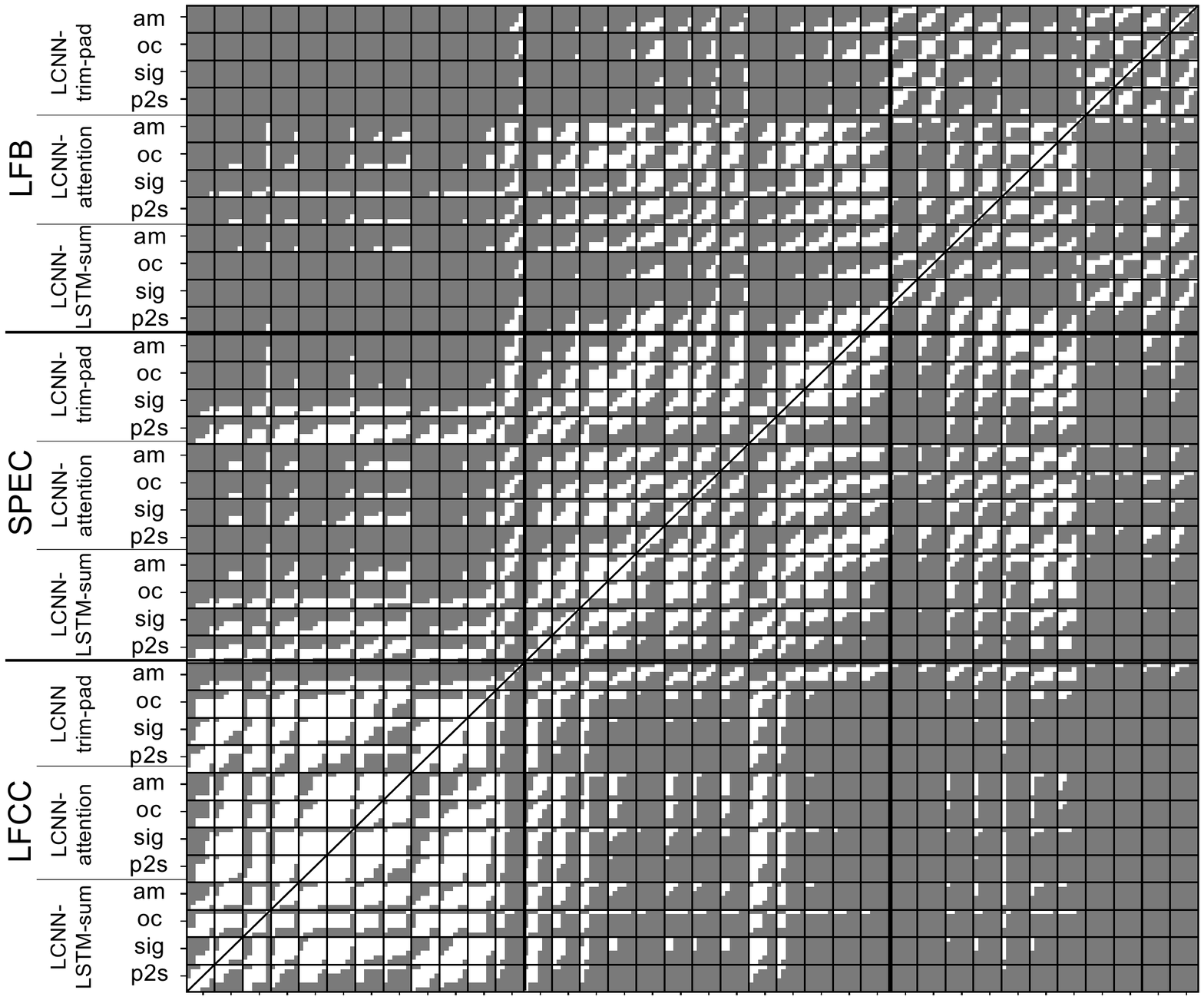}
\vspace{-8mm}
\caption{Statistical significance test on EERs \cite{bengio2004statistical}, given $\alpha=0.05$ with Holm-Bonferroni correction. Significant difference is indicated by dark grey, otherwise by white. Each square in the black frames contains $6\times6$ entries and denotes pair-wise tests between six training-evaluation rounds of two models. The six rounds of each model were in the same order as that in Table~\ref{tab:eer}.}
\label{fig:sig_test}
\end{figure}

Table~\ref{tab:eer} lists the EERs and min t-DCFs on the evaluation set. 
The outcomes of the pair-wise statistical analysis are plotted in Fig.~\ref{fig:sig_test}, where white and dark grey indicate insignificant and significant statistical differences, respectively.

We observed that the same model may perform quite differently when simply changing the random seed, and this intra-model difference can be statistically significant, as the blocks on the anti-diagonal line of Fig.~\ref{fig:sig_test} demonstrate. 
Although such variation was expected for neural-network-based CMs, the difference between the best and worst runs of the same model may be larger than the inter-model differences. This finding calls for caution when comparing and interpreting the EERs of different models.
Particularly, the commonly used baseline model LFCC \systemI\ achieved a low EER of 2.54\% when it was well initialized, which is a stronger baseline than those in many other studies. 

Despite the variation,  \systemIII\ with the LFCC and P2S achieved EERs in the range of (1.92\%, 3.10\%), which is in the same tier as the lowest EERs reported in other studies without data augmentation, e.g., 2.19\% in \cite{zhang2020one}, 3.49\% in \cite{Chen2020Odyssey}, and 4.04\% in \cite{Nautsch2021}. 
The best single run with 1.92\% EER is currently the lowest EER value on the database without data augmentation as far as we know, and this EER is statistically significantly different from most of the others, as the bottom row of Fig.~\ref{fig:sig_test} illustrates.
Some techniques used by the model may be promising for this task.

One of the techniques is the new P2SGrad-based loss function. 
With the commonly used LFCC front end, the min and max EERs of the P2SGrad-based models were lower than those using other loss functions for the three tested network structures. Furthermore, the P2SGrad-based loss had no hyper-parameter. 
Note that the AM and OC-softmax used the best hyper-parameter configuration reported in the literature \cite{zhang2020one,Chen2020Odyssey}. 


Among the three network structures,  \systemI 's EER was comparable to the other two when using the LFCC and sigmoid or P2SGrad. 
However, while \systemII\ and \systemIII\ had roughly $190 \pm 30$k and $290 \pm 30$k parameters, respectively\footnote{Parameter size varies due to additional FC layers for different loss functions and spectrogram compression (Sec.~\ref{seq:exp_models}).}, \systemI\ had more than $860$k. Furthermore, the single FC layer after a flattening operation in \systemI\ took $710$k parameters, which was above 80\% of the network size. 
This monolithic layer conducted $\mathbb{R}^{\frac{K}{L}D_h}\rightarrow\mathbb{R}$, and its parameter size was proportional to the fixed length $K$ of the input features. 
Such a configuration is inefficient for speech anti-spoofing.

Finally, we observed that the LFCC leads to better performance than the other two frond ends on the database, and the difference was statistically significant as Fig.~\ref{fig:sig_test} shows. Except \systemI\ with the AM-softmax, all models with the LFCC had an EER below 4.64\%, which is comparable to the best LFCC-based single system in ASVspoof2019 \cite{Nautsch2021}. 
Although LFB has demonstrated good performance on ResNet in another study \cite{Chen2020Odyssey}, it may not be the best choice for LCNN back ends. This possibility will be explored in future work.

\section{Conclusion}
We compared commonly used components for neural-network-based CMs on the ASVspoof2019 LA database. 
By training and evaluating each CM for multiple runs, we observed that the intra-model differences due to random initial seeds may be statistically significant. 
We tentatively recommend similar analysis or multiple training-evaluation rounds on this database.

Despite the variation, our results suggest a few promising components for CMs.
On neural networks, although CNNs with fixed-size input are widely used, our results suggest that networks with attention or even average-based pooling are potentially more efficient to process varied-length input trials for speech anti-spoofing.  
On loss functions, the simple sigmoid function is comparable to margin-based softmax for LCNN-based CMs, and a newly proposed loss function with P2SGrad also performed decently through six rounds of training and evaluation using the LFCC front end.
With the best combination of LFCC front end, LCNN-LSTM back end, and P2SGrad-based loss function, the lowest EER on the ASVspoof2019 LA evaluation set reached 1.92\% in this study.  We released the code and trained models for reproducible research.

\section{Acknowledgements}
This study is supported by JST CREST Grants (JPMJCR18A6 and JPMJCR20D3), MEXT KAKENHI Grants (16H06302, 18H04120, 18H04112, and 18KT0051), Japan, and Google AI for Japan program. Some of the experiments were conducted on TSUBAME 3.0 of Tokyo Institute of Technology. We thank other ASVspoof2019 organizers and the reviewers for the comments. Some replies are added to the appendix on Arxiv.

\clearpage
\newpage
\bibliographystyle{IEEEtran}
\bibliography{../mybib}


\clearpage
\newpage
\appendix
\onecolumn
\section{Appendix}

\subsection{Decomposed EER and min-tDCFs}
\begin{figure}[h]
\centering
\includegraphics[width=0.8\textwidth]{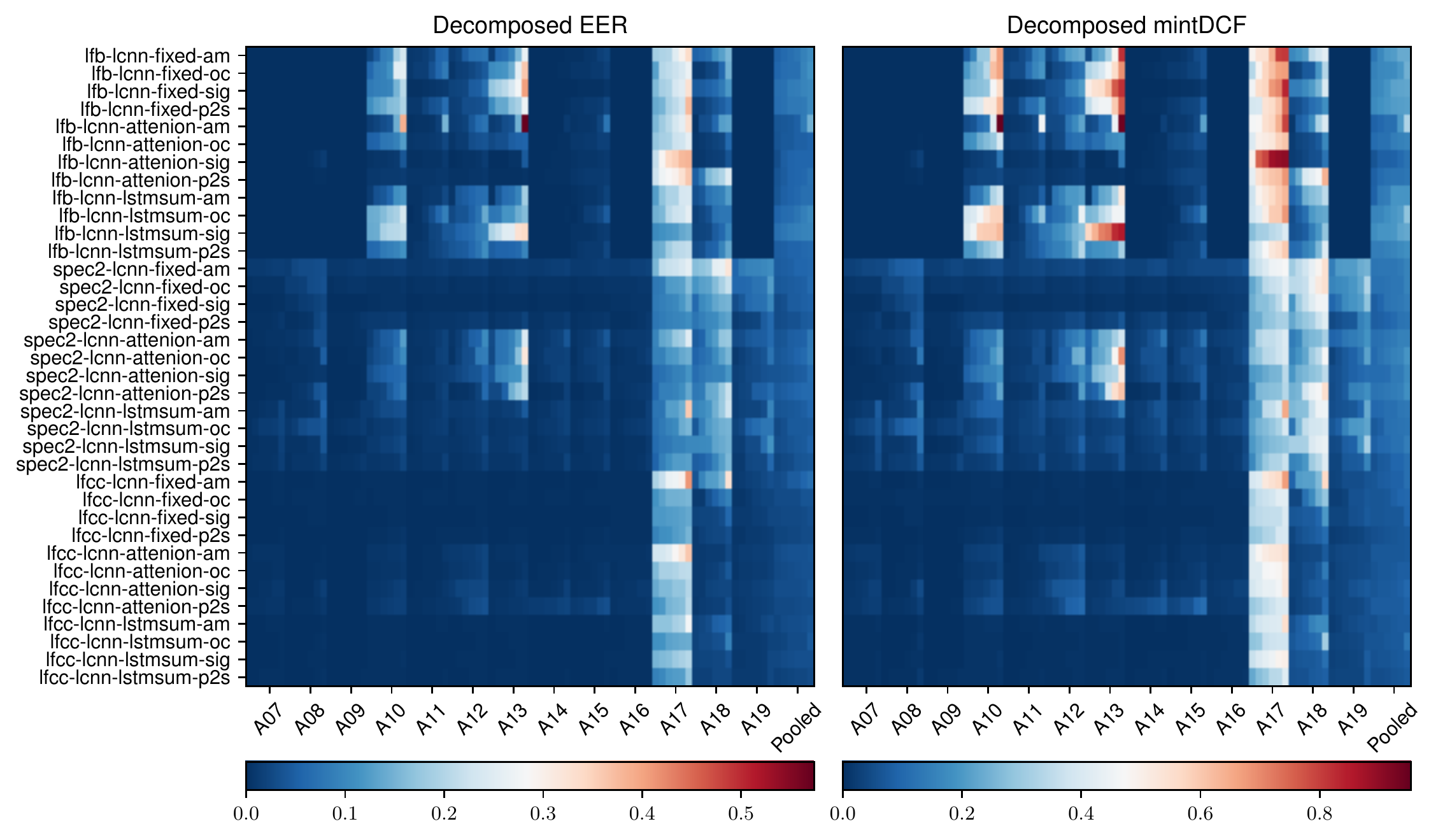}
\caption{Decomposed and pooled EER and mint t-DCFs. Each row has $14\times 6$ entries, where $14$ denotes $13$ attacks and one pooled result, and where $6$ denotes the six training-evaluation rounds on each attack. }
\label{fig:app_eers}
\end{figure}

Decomposed EER and min tDCFs are plotted in Figure~\ref{fig:app_eers}.
Obviously, A17 is the most difficult attack for models compared in this study.

\subsection{Additional experiment results}
\subsubsection{On spectrogram front end}
We mentioned in Sec.~\ref{seq:exp_models} that the models using the spectrogram front end has an additional FC layer that transforms the input spectrogram of shape ${N\times 257}$ into a hidden feature map ${N\times 60}$, where $N$ is the frame number. Without this layer, the performance severely degraded, as Tab.~\ref{tab:app_eer} demonstrates.

\begin{table*}[h]
\caption{Effect of using a FC layer to pre-process spectrograms. EER and min t-DCF on ASVspoof2019 LA evaluation set}
\vspace{-6mm}
\begin{center}
\resizebox{\textwidth}{!}{
\setlength{\tabcolsep}{2pt}
\begin{tabular}{cc|cccccc|cccccc|cccccc|cccccc}
\multicolumn{26}{c}{EERs}   \\
\toprule
  \multicolumn{2}{c}{} &  \multicolumn{6}{c}{AM-softmax} &  \multicolumn{6}{c}{OC-softmax} &  \multicolumn{6}{c}{Sigmoid} &  \multicolumn{6}{c}{MSE for P2SGrad}   \\
  \cmidrule(r){3-26}
\multicolumn{2}{c}{}   &  I  &  II  &  III   &  IV   &  V   &  \multicolumn{1}{c}{VI}   & I  &  II  &  III   &  IV   &  V   &   \multicolumn{1}{c}{VI}      & I  &  II  &  III   &  IV   &  V   &   \multicolumn{1}{c}{VI}       &  I  &  II  &  III   &  IV   &  V   &   \multicolumn{1}{c}{VI}   \\ 
\midrule
\multirow{2}{*}{W/o FC layer} &   \systemII    & \cellcolor[rgb]{0.74, 0.74, 0.74} 29.61 & \cellcolor[rgb]{0.73, 0.73, 0.73} 30.30 & \cellcolor[rgb]{0.70, 0.70, 0.70} 32.96 & \cellcolor[rgb]{0.70, 0.70, 0.70} 33.04 & \cellcolor[rgb]{0.62, 0.62, 0.62} 39.20 & \cellcolor[rgb]{0.60, 0.60, 0.60} 41.25 & \cellcolor[rgb]{0.76, 0.76, 0.76} 27.41 & \cellcolor[rgb]{0.75, 0.75, 0.75} 28.72 & \cellcolor[rgb]{0.74, 0.74, 0.74} 29.42 & \cellcolor[rgb]{0.58, 0.58, 0.58} 42.56 & \cellcolor[rgb]{0.58, 0.58, 0.58} 43.28 & \cellcolor[rgb]{0.57, 0.57, 0.57} 43.78 & \cellcolor[rgb]{0.77, 0.77, 0.77} 26.00 & \cellcolor[rgb]{0.77, 0.77, 0.77} 26.10 & \cellcolor[rgb]{0.72, 0.72, 0.72} 31.37 & \cellcolor[rgb]{0.66, 0.66, 0.66} 35.65 & \cellcolor[rgb]{0.65, 0.65, 0.65} 36.48 & \cellcolor[rgb]{0.61, 0.61, 0.61} 39.90 & \cellcolor[rgb]{0.80, 0.80, 0.80} 23.83 & \cellcolor[rgb]{0.79, 0.79, 0.79} 24.26 & \cellcolor[rgb]{0.76, 0.76, 0.76} 27.21 & \cellcolor[rgb]{0.73, 0.73, 0.73} 30.55 & \cellcolor[rgb]{0.70, 0.70, 0.70} 32.93 & \cellcolor[rgb]{0.63, 0.63, 0.63} 38.30\\ 
 &  \systemIII     & \cellcolor[rgb]{0.92, 0.92, 0.92} 10.81 & \cellcolor[rgb]{0.90, 0.90, 0.90} 13.28 & \cellcolor[rgb]{0.79, 0.79, 0.79} 24.65 & \cellcolor[rgb]{0.77, 0.77, 0.77} 26.01 & \cellcolor[rgb]{0.76, 0.76, 0.76} 27.80 & \cellcolor[rgb]{0.75, 0.75, 0.75} 28.21 & \cellcolor[rgb]{0.95, 0.95, 0.95} 7.83 & \cellcolor[rgb]{0.94, 0.94, 0.94} 8.80 & \cellcolor[rgb]{0.92, 0.92, 0.92} 11.57 & \cellcolor[rgb]{0.90, 0.90, 0.90} 13.23 & \cellcolor[rgb]{0.90, 0.90, 0.90} 13.57 & \cellcolor[rgb]{0.88, 0.88, 0.88} 15.68 & \cellcolor[rgb]{0.96, 0.96, 0.96} 6.90 & \cellcolor[rgb]{0.95, 0.95, 0.95} 8.55 & \cellcolor[rgb]{0.87, 0.87, 0.87} 16.72 & \cellcolor[rgb]{0.84, 0.84, 0.84} 19.95 & \cellcolor[rgb]{0.79, 0.79, 0.79} 24.31 & \cellcolor[rgb]{0.66, 0.66, 0.66} 36.22 & \cellcolor[rgb]{0.95, 0.95, 0.95} 8.25 & \cellcolor[rgb]{0.90, 0.90, 0.90} 13.17 & \cellcolor[rgb]{0.82, 0.82, 0.82} 21.24 & \cellcolor[rgb]{0.82, 0.82, 0.82} 21.70 & \cellcolor[rgb]{0.55, 0.55, 0.55} 45.74 & \cellcolor[rgb]{0.37, 0.37, 0.37} 69.77\\ 
 \midrule
\multirow{2}{*}{W/ FC layer} &  \systemII     & \cellcolor[rgb]{0.98, 0.98, 0.98} 4.02 & \cellcolor[rgb]{0.98, 0.98, 0.98} 4.08 & \cellcolor[rgb]{0.97, 0.97, 0.97} 4.99 & \cellcolor[rgb]{0.97, 0.97, 0.97} 5.22 & \cellcolor[rgb]{0.97, 0.97, 0.97} 5.57 & \cellcolor[rgb]{0.95, 0.95, 0.95} 8.20 & \cellcolor[rgb]{0.98, 0.98, 0.98} 4.05 & \cellcolor[rgb]{0.98, 0.98, 0.98} 4.55 & \cellcolor[rgb]{0.97, 0.97, 0.97} 4.94 & \cellcolor[rgb]{0.97, 0.97, 0.97} 5.67 & \cellcolor[rgb]{0.96, 0.96, 0.96} 6.31 & \cellcolor[rgb]{0.95, 0.95, 0.95} 8.47 & \cellcolor[rgb]{0.98, 0.98, 0.98} 3.92 & \cellcolor[rgb]{0.98, 0.98, 0.98} 4.04 & \cellcolor[rgb]{0.98, 0.98, 0.98} 4.42 & \cellcolor[rgb]{0.97, 0.97, 0.97} 4.91 & \cellcolor[rgb]{0.97, 0.97, 0.97} 4.95 & \cellcolor[rgb]{0.96, 0.96, 0.96} 6.62 & \cellcolor[rgb]{0.98, 0.98, 0.98} 4.70 & \cellcolor[rgb]{0.97, 0.97, 0.97} 5.03 & \cellcolor[rgb]{0.97, 0.97, 0.97} 5.60 & \cellcolor[rgb]{0.97, 0.97, 0.97} 5.88 & \cellcolor[rgb]{0.96, 0.96, 0.96} 6.35 & \cellcolor[rgb]{0.96, 0.96, 0.96} 6.50\\ 
 & \systemIII      & \cellcolor[rgb]{0.98, 0.98, 0.98} 3.96 & \cellcolor[rgb]{0.98, 0.98, 0.98} 4.04 & \cellcolor[rgb]{0.98, 0.98, 0.98} 4.38 & \cellcolor[rgb]{0.98, 0.98, 0.98} 4.52 & \cellcolor[rgb]{0.97, 0.97, 0.97} 5.13 & \cellcolor[rgb]{0.96, 0.96, 0.96} 5.97 & \cellcolor[rgb]{0.99, 0.99, 0.99} 2.81 & \cellcolor[rgb]{0.99, 0.99, 0.99} 3.41 & \cellcolor[rgb]{0.98, 0.98, 0.98} 4.49 & \cellcolor[rgb]{0.98, 0.98, 0.98} 4.50 & \cellcolor[rgb]{0.98, 0.98, 0.98} 4.65 & \cellcolor[rgb]{0.97, 0.97, 0.97} 4.91 & \cellcolor[rgb]{0.99, 0.99, 0.99} 3.29 & \cellcolor[rgb]{0.99, 0.99, 0.99} 3.56 & \cellcolor[rgb]{0.98, 0.98, 0.98} 3.82 & \cellcolor[rgb]{0.98, 0.98, 0.98} 4.45 & \cellcolor[rgb]{0.98, 0.98, 0.98} 4.61 & \cellcolor[rgb]{0.97, 0.97, 0.97} 5.44 & \cellcolor[rgb]{1.00, 1.00, 1.00} 2.37 & \cellcolor[rgb]{0.99, 0.99, 0.99} 2.91 & \cellcolor[rgb]{0.99, 0.99, 0.99} 3.00 & \cellcolor[rgb]{0.98, 0.98, 0.98} 3.94 & \cellcolor[rgb]{0.98, 0.98, 0.98} 4.26 & \cellcolor[rgb]{0.98, 0.98, 0.98} 4.37\\ 
 \bottomrule
 \\
\multicolumn{26}{c}{min t-DCF}   \\
\toprule
 \multirow{2}{*}{W/o FC layer} &  \systemII     & \cellcolor[rgb]{0.66, 0.66, 0.66} 0.532 & \cellcolor[rgb]{0.65, 0.65, 0.65} 0.542 & \cellcolor[rgb]{0.57, 0.57, 0.57} 0.637 & \cellcolor[rgb]{0.48, 0.48, 0.48} 0.776 & \cellcolor[rgb]{0.50, 0.50, 0.50} 0.741 & \cellcolor[rgb]{0.49, 0.49, 0.49} 0.763 & \cellcolor[rgb]{0.67, 0.67, 0.67} 0.517 & \cellcolor[rgb]{0.56, 0.56, 0.56} 0.651 & \cellcolor[rgb]{0.56, 0.56, 0.56} 0.658 & \cellcolor[rgb]{0.52, 0.52, 0.52} 0.707 & \cellcolor[rgb]{0.52, 0.52, 0.52} 0.707 & \cellcolor[rgb]{0.48, 0.48, 0.48} 0.773 & \cellcolor[rgb]{0.68, 0.68, 0.68} 0.504 & \cellcolor[rgb]{0.62, 0.62, 0.62} 0.576 & \cellcolor[rgb]{0.68, 0.68, 0.68} 0.505 & \cellcolor[rgb]{0.55, 0.55, 0.55} 0.671 & \cellcolor[rgb]{0.48, 0.48, 0.48} 0.767 & \cellcolor[rgb]{0.52, 0.52, 0.52} 0.719 & \cellcolor[rgb]{0.72, 0.72, 0.72} 0.455 & \cellcolor[rgb]{0.72, 0.72, 0.72} 0.461 & \cellcolor[rgb]{0.72, 0.72, 0.72} 0.458 & \cellcolor[rgb]{0.67, 0.67, 0.67} 0.512 & \cellcolor[rgb]{0.69, 0.69, 0.69} 0.500 & \cellcolor[rgb]{0.64, 0.64, 0.64} 0.553\\ 
 & \systemIII      & \cellcolor[rgb]{0.87, 0.87, 0.87} 0.250 & \cellcolor[rgb]{0.79, 0.79, 0.79} 0.365 & \cellcolor[rgb]{0.73, 0.73, 0.73} 0.448 & \cellcolor[rgb]{0.74, 0.74, 0.74} 0.442 & \cellcolor[rgb]{0.71, 0.71, 0.71} 0.474 & \cellcolor[rgb]{0.69, 0.69, 0.69} 0.500 & \cellcolor[rgb]{0.92, 0.92, 0.92} 0.191 & \cellcolor[rgb]{0.90, 0.90, 0.90} 0.210 & \cellcolor[rgb]{0.87, 0.87, 0.87} 0.255 & \cellcolor[rgb]{0.80, 0.80, 0.80} 0.354 & \cellcolor[rgb]{0.84, 0.84, 0.84} 0.301 & \cellcolor[rgb]{0.76, 0.76, 0.76} 0.404 & \cellcolor[rgb]{0.92, 0.92, 0.92} 0.191 & \cellcolor[rgb]{0.87, 0.87, 0.87} 0.252 & \cellcolor[rgb]{0.81, 0.81, 0.81} 0.337 & \cellcolor[rgb]{0.64, 0.64, 0.64} 0.547 & \cellcolor[rgb]{0.73, 0.73, 0.73} 0.452 & \cellcolor[rgb]{0.64, 0.64, 0.64} 0.554 & \cellcolor[rgb]{0.91, 0.91, 0.91} 0.200 & \cellcolor[rgb]{0.87, 0.87, 0.87} 0.258 & \cellcolor[rgb]{0.80, 0.80, 0.80} 0.356 & \cellcolor[rgb]{0.79, 0.79, 0.79} 0.363 & \cellcolor[rgb]{0.62, 0.62, 0.62} 0.572 & \cellcolor[rgb]{0.37, 0.37, 0.37} 1.000\\ 
  \midrule
 \multirow{2}{*}{W/ FC layer} &    \systemII   & \cellcolor[rgb]{0.96, 0.96, 0.96} 0.119 & \cellcolor[rgb]{0.96, 0.96, 0.96} 0.117 & \cellcolor[rgb]{0.95, 0.95, 0.95} 0.131 & \cellcolor[rgb]{0.94, 0.94, 0.94} 0.153 & \cellcolor[rgb]{0.95, 0.95, 0.95} 0.139 & \cellcolor[rgb]{0.94, 0.94, 0.94} 0.156 & \cellcolor[rgb]{0.97, 0.97, 0.97} 0.105 & \cellcolor[rgb]{0.96, 0.96, 0.96} 0.126 & \cellcolor[rgb]{0.95, 0.95, 0.95} 0.141 & \cellcolor[rgb]{0.94, 0.94, 0.94} 0.153 & \cellcolor[rgb]{0.94, 0.94, 0.94} 0.162 & \cellcolor[rgb]{0.91, 0.91, 0.91} 0.194 & \cellcolor[rgb]{0.97, 0.97, 0.97} 0.109 & \cellcolor[rgb]{0.97, 0.97, 0.97} 0.109 & \cellcolor[rgb]{0.96, 0.96, 0.96} 0.115 & \cellcolor[rgb]{0.95, 0.95, 0.95} 0.140 & \cellcolor[rgb]{0.95, 0.95, 0.95} 0.141 & \cellcolor[rgb]{0.93, 0.93, 0.93} 0.173 & \cellcolor[rgb]{0.95, 0.95, 0.95} 0.135 & \cellcolor[rgb]{0.96, 0.96, 0.96} 0.125 & \cellcolor[rgb]{0.95, 0.95, 0.95} 0.143 & \cellcolor[rgb]{0.96, 0.96, 0.96} 0.129 & \cellcolor[rgb]{0.94, 0.94, 0.94} 0.153 & \cellcolor[rgb]{0.92, 0.92, 0.92} 0.183\\ 
 &   \systemIII    & \cellcolor[rgb]{0.97, 0.97, 0.97} 0.106 & \cellcolor[rgb]{0.97, 0.97, 0.97} 0.102 & \cellcolor[rgb]{0.96, 0.96, 0.96} 0.116 & \cellcolor[rgb]{0.96, 0.96, 0.96} 0.118 & \cellcolor[rgb]{0.95, 0.95, 0.95} 0.136 & \cellcolor[rgb]{0.95, 0.95, 0.95} 0.134 & \cellcolor[rgb]{0.99, 0.99, 0.99} 0.077 & \cellcolor[rgb]{0.98, 0.98, 0.98} 0.088 & \cellcolor[rgb]{0.96, 0.96, 0.96} 0.118 & \cellcolor[rgb]{0.96, 0.96, 0.96} 0.130 & \cellcolor[rgb]{0.96, 0.96, 0.96} 0.121 & \cellcolor[rgb]{0.96, 0.96, 0.96} 0.126 & \cellcolor[rgb]{0.98, 0.98, 0.98} 0.087 & \cellcolor[rgb]{0.97, 0.97, 0.97} 0.105 & \cellcolor[rgb]{0.97, 0.97, 0.97} 0.102 & \cellcolor[rgb]{0.96, 0.96, 0.96} 0.122 & \cellcolor[rgb]{0.96, 0.96, 0.96} 0.122 & \cellcolor[rgb]{0.95, 0.95, 0.95} 0.146 & \cellcolor[rgb]{1.00, 1.00, 1.00} 0.060 & \cellcolor[rgb]{0.99, 0.99, 0.99} 0.077 & \cellcolor[rgb]{0.99, 0.99, 0.99} 0.079 & \cellcolor[rgb]{0.97, 0.97, 0.97} 0.097 & \cellcolor[rgb]{0.97, 0.97, 0.97} 0.097 & \cellcolor[rgb]{0.96, 0.96, 0.96} 0.117\\ 
  \bottomrule
\end{tabular}
}
\end{center}
\vspace{-6mm}
\label{tab:app_eer}
\end{table*}%

\subsubsection{LSTM with Attention-based pooling}
We did experiment to check whether the attention-based pooling can be stacked on the top of LSTM. As the last row of Tab.~\ref{tab:app_eer_att} shows, the improvement is trivial, and thus we did not include this model for further experiment.

\begin{table}[h]
\caption{Effect of adding attention-based pooling after LSTM. Only EERs are listed.}
\centering
\begin{tabular}{cccccccc}
  \toprule
 &      &  I  &  II  &  III   &  IV   &  V  & VI  \\ 
 \midrule
 &  \systemII     & \cellcolor[rgb]{0.66, 0.66, 0.66} 2.72 & \cellcolor[rgb]{0.61, 0.61, 0.61} 2.81 & \cellcolor[rgb]{0.57, 0.57, 0.57} 2.91 & \cellcolor[rgb]{0.52, 0.52, 0.52} 3.03 & \cellcolor[rgb]{0.45, 0.45, 0.45} 3.22 & \cellcolor[rgb]{0.42, 0.42, 0.42} 3.30\\ 
 &  \systemIII     & \cellcolor[rgb]{1.00, 1.00, 1.00} 1.92 & \cellcolor[rgb]{0.94, 0.94, 0.94} 2.09 & \cellcolor[rgb]{0.80, 0.80, 0.80} 2.43 & \cellcolor[rgb]{0.77, 0.77, 0.77} 2.50 & \cellcolor[rgb]{0.71, 0.71, 0.71} 2.62 & \cellcolor[rgb]{0.49, 0.49, 0.49} 3.10\\ 
 &  \systemIII + attention-based pooling     & \cellcolor[rgb]{0.94, 0.94, 0.94} 2.08 & \cellcolor[rgb]{0.89, 0.89, 0.89} 2.20 & \cellcolor[rgb]{0.88, 0.88, 0.88} 2.23 & \cellcolor[rgb]{0.80, 0.80, 0.80} 2.43 & \cellcolor[rgb]{0.60, 0.60, 0.60} 2.84 & \cellcolor[rgb]{0.37, 0.37, 0.37} 3.52\\ 
   \bottomrule
\end{tabular}
\label{tab:app_eer_att}
\end{table}

One of reviewer asked why we added Bi-LSTM to the experimental model. One reason is that the convolution in CNN only covers a fixed receptive field. For example, for a CNN with a single convolution layer and kernel size of 3, it only computes an output value using input features within the kernel. Although the receptive field can be expanded by stacking CNN layers and increasing the dilation size, it is fixed and may not cover the whole input trial. In contrast, Bi-LSTM process all the input feature frames, and each output feature from the Bi-LSTM depends on all the input feature frames.

\subsection{Fusion}
Here we fuse the models by averaging the scores.

\subsubsection{Fuse homogenuous model from different training rounds}
Since each model was trained for six rounds, we may fuse the trained models from the six sounds. 
However, the performance gain is trivial. One possible reason is that these sub-models are not every different from each other: they may look for similar patterns from the input features. This is also indicated by the decomposed EERs in Figure 4, where sub-models performed similarly on different attackers.

\begin{figure}[h]
\centering
      \begin{subfigure}[b]{0.45\textwidth}
         \centering
         \includegraphics[width=0.8\textwidth]{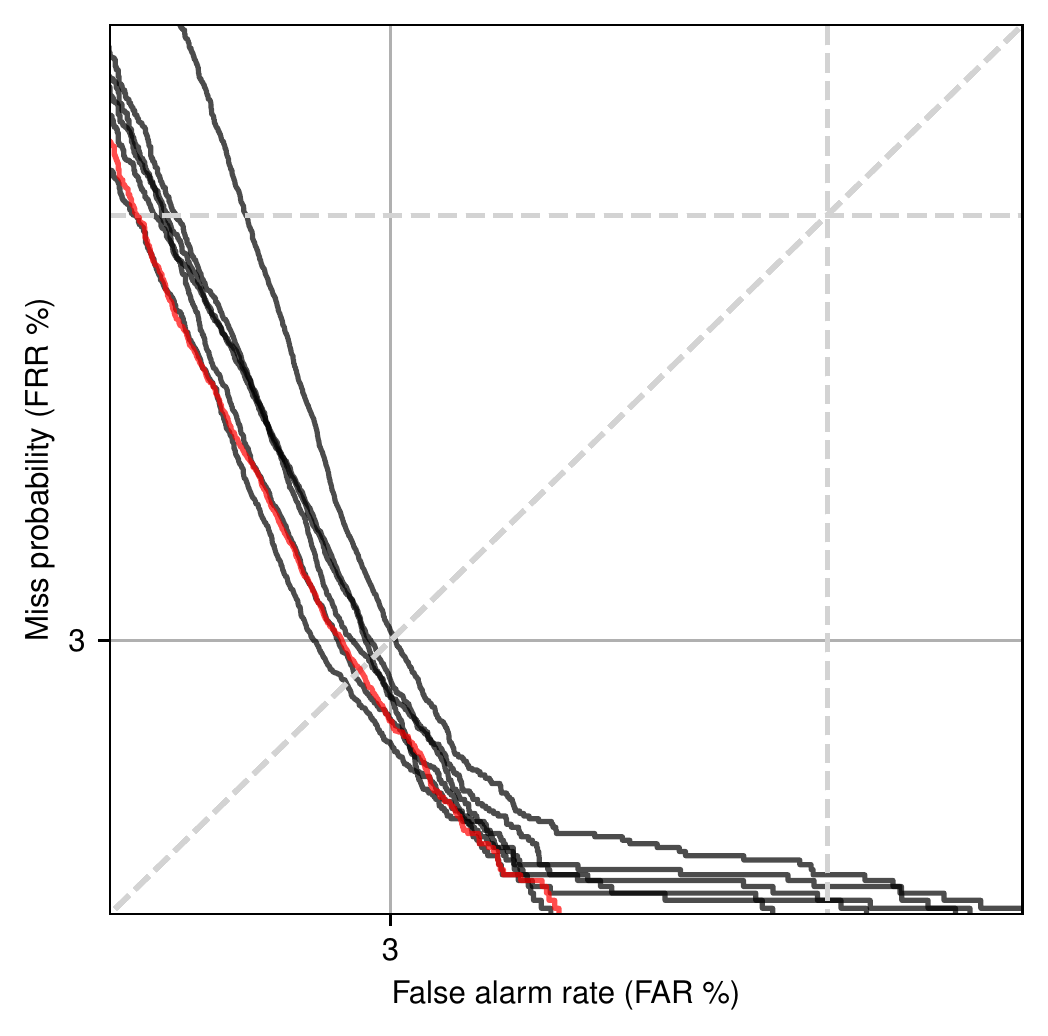}
         \caption{Fuse LFCC \systemIII\ with P2SGrad (I to VI). \textcolor{red}{EER of fused model is  2.828\%}.}
         \label{fig:fus_p2s}
     \end{subfigure}
     \hfill
      \begin{subfigure}[b]{0.45\textwidth}
         \centering
         \includegraphics[width=0.8\textwidth]{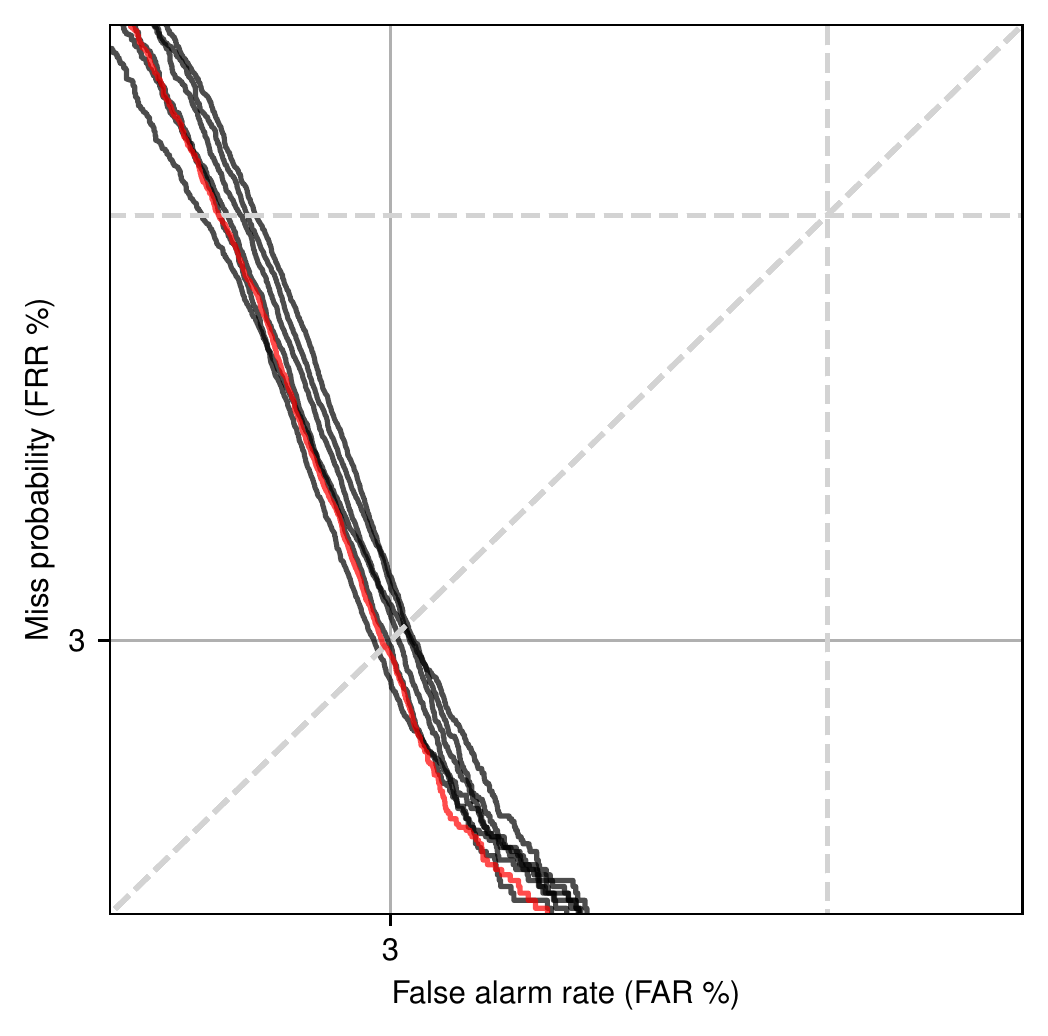}
         \caption{Fuse LFCC \systemIII\ with Sigmoid (I to VI). \textcolor{red}{EER of fused model is  2.216\%}.}
         \label{fig:fus_sig}
     \end{subfigure}
     \hfill
\caption{Fusion of homogeneous models from the six training rounds. \textcolor{red}{DET curve in red color} marks the fused model while black curves are sub-models.}
\label{fig:fus1}
\end{figure}

\subsubsection{Fuse different models}
We may fuse models using different front-ends, network architectures, and training criteria. 
Fusion models with different front-ends achieved the lowest EER 1.074\%, which is one of the lowest primary system EER on ASVspoof2019 LA. Fusion over network types or training criteria is less effective. 

\begin{figure}[h]
\centering
      \begin{subfigure}[b]{0.45\textwidth}
         \centering
         \includegraphics[width=0.8\textwidth]{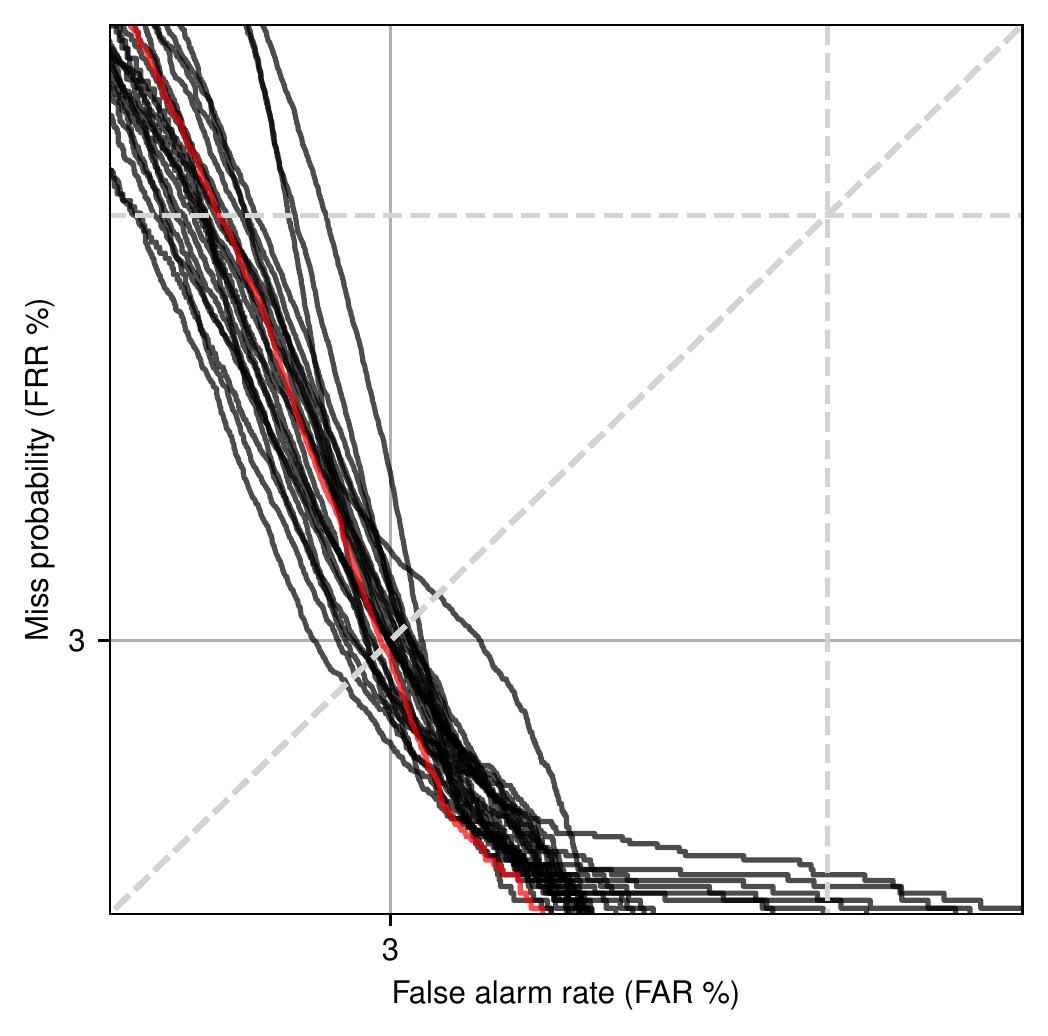}
         \caption{Fusion over training criteria: LFCC \systemIII\ using Sigmoid, P2S, AM, and OC-softmax. \textcolor{red}{EER of fused model is  2.83\%}.}
         \label{fig:fus_cri}
     \end{subfigure}
     \hfill
      \begin{subfigure}[b]{0.45\textwidth}
         \centering
         \includegraphics[width=0.8\textwidth]{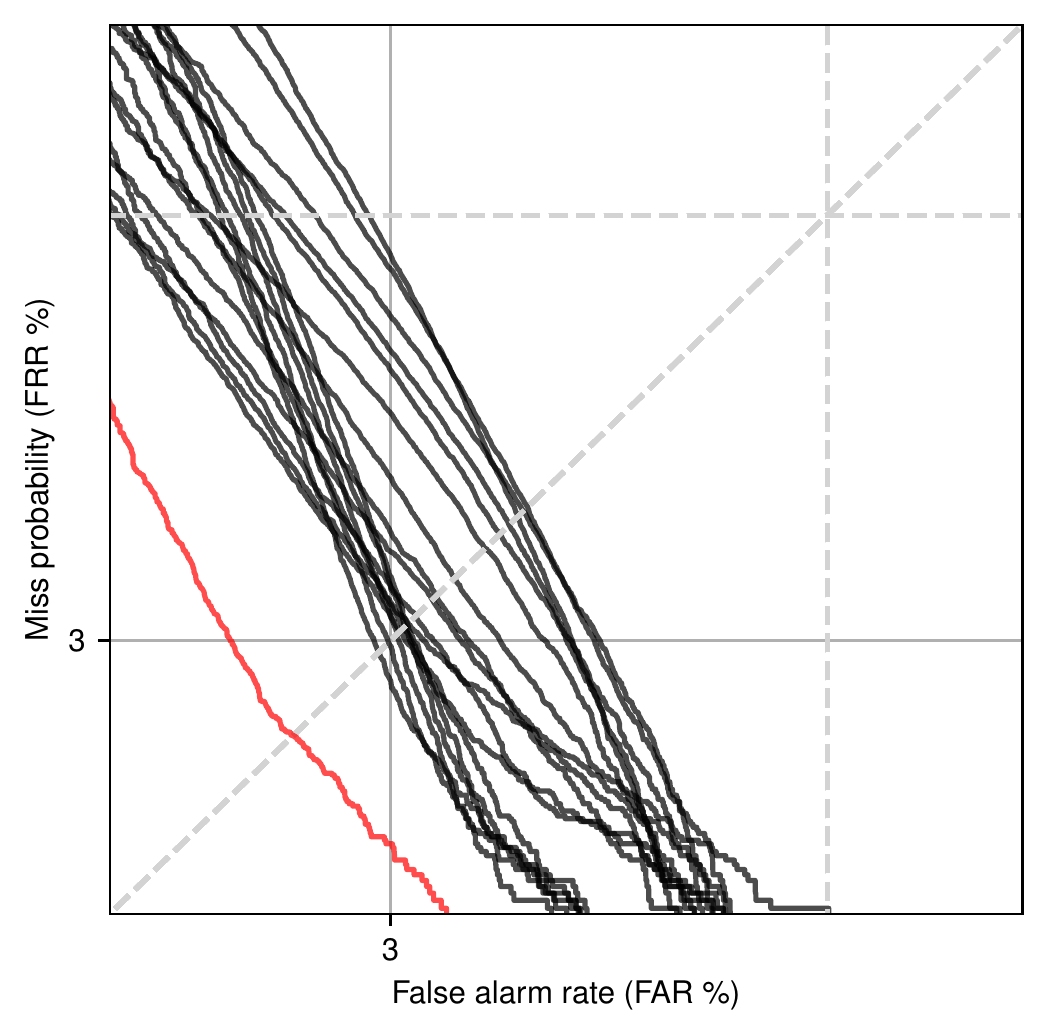}
         \caption{Fuse over front-ends, \systemIII\ Sigmoid using LFCC, LFB, and, Spectra. \textcolor{red}{EER of fused model is  1.074\%}.}
         \label{fig:fus_front}
     \end{subfigure}
     \hfill
     
     \begin{subfigure}[b]{0.45\textwidth}
         \centering
         \includegraphics[width=0.8\textwidth]{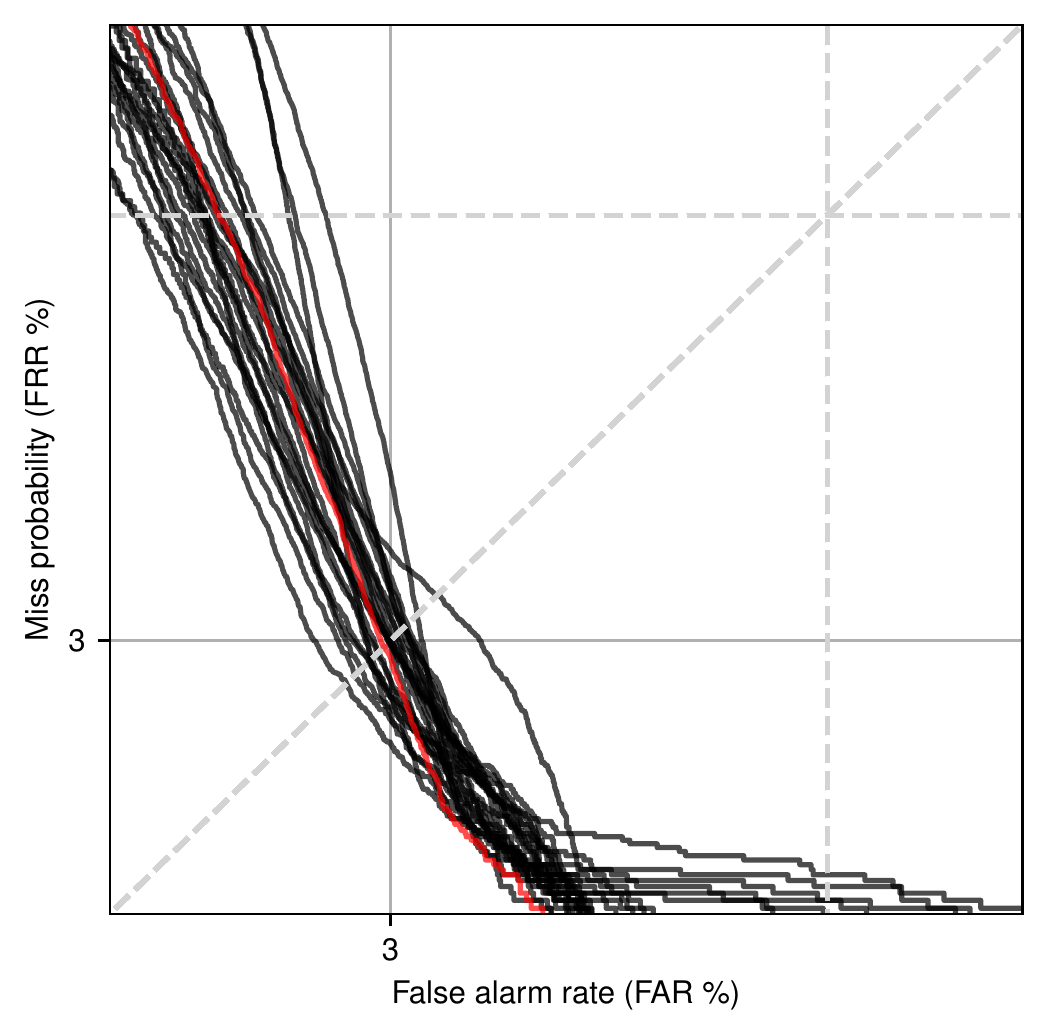}
         \caption{Fuse over network types: \systemI, \systemII, \systemIII, using LFCC Sigmoid (including models from I to VI training rounds). \textcolor{red}{EER of fused model is  2.83\%}.}
         \label{fig:fus_net}
     \end{subfigure}
     \hfill
\caption{Fusion of different models. \textcolor{red}{DET curve in red color} marks the fused model while black curves are sub-models.}
\label{fig:fus1}
\end{figure}

\end{document}